\documentclass[journal ]{new-aiaa}
\usepackage[utf8]{inputenc}
\usepackage{textcomp}

\usepackage{graphicx}
\usepackage{amsmath}
\usepackage[version=4]{mhchem}
\usepackage{siunitx}
\usepackage{longtable,tabularx}
\usepackage{soul}
\usepackage{subcaption}
\usepackage{bm}

\title{Regression Trees on Grassmann Manifold \\ for Adapting Reduced-Order Models}

\author{X. Liu and X.C. Liu}
\affil{Department of Industrial Engineering, University of Arkansas, Fayetteville, AR, 72701}

\begin{document}

\maketitle

\begin{abstract}
Low dimensional and computationally less expensive Reduced-Order Models (ROMs) have been widely used to capture the dominant behaviors of high-dimensional systems. A ROM can be obtained, using the well-known Proper Orthogonal Decomposition (POD), by projecting the full-order model to a subspace spanned by modal basis modes which are learned from experimental, simulated or observational data, i.e., training data. However, the optimal basis can change with the parameter settings. When a ROM, constructed using the POD basis obtained from training data, is applied to new parameter settings, the model often lacks robustness against the change of parameters in design, control, and other real-time operation problems. This paper proposes to use regression trees on Grassmann Manifold to learn the mapping between parameters and  POD bases that span the low-dimensional subspaces onto which full-order models are projected. Motivated by the fact that a subspace spanned by a POD basis can be viewed as a point in the Grassmann manifold, we propose to grow a tree by repeatedly splitting the tree node to maximize the Riemannian distance between the two subspaces spanned by the predicted POD bases on the left and right daughter nodes. Five numerical examples are presented to comprehensively demonstrate the performance of the proposed method, and compare the proposed tree-based method to the existing interpolation method for POD basis and the use of global POD basis. The results show that the proposed tree-based method is capable of establishing the mapping between parameters and POD bases, and thus adapt ROMs for new parameters. 
\end{abstract}

\section*{Nomenclature}

{\renewcommand\arraystretch{1.0}
\noindent\begin{longtable*}{@{}l @{\quad=\quad} l@{}}
$\Lambda$  & a set of parameters \\
$\bm{\lambda}_i$ &  a vector representing the $i$-th parameter settings \\
$\bm{D}(\bm{\lambda})$ & a snapshot data matrix obtained under parameter setting $\bm{\lambda}$ \\
$x(t, \bm{s}; \bm{\lambda})$ & the solution of the full-order system given the parameter  $\bm{\lambda}$ \\
$\bm{\Phi}$ & the matrix basis \\
$\mathcal{ST}(r,n)$ & Stiefel manifold\\
$\mathcal{G}(r,n)$ & Grassmann manifold\\
$\sigma$ & singular value \\

\end{longtable*}}

\section{Introduction} \label{sec:one}
\lettrine{T}{he} complex dynamics of many engineering, physical and biological systems are often described by Partial Differential Equations (PDEs). Because solving these potentially high-dimensional PDEs can be computationally expensive, 
low-dimensional Reduced-Order Models (ROMs) have been widely utilized to capture the dominant behaviors of the original systems. This idea is primarily motivated and justified by the ubiquitous observations in engineering and science that there often exist low-dimensional patterns embedded in high-dimensional systems.  

A ROM can be obtained by projecting the full-order model to a subspace spanned by chosen modal basis modes. For example, based on experimental or simulated data generated from full-order models under predetermined parameter settings, the Proper Orthogonal Decomposition (POD) can be used to generate the optimal basis modes in an $L_2$ sense. However, the optimal bases can change with parameter settings. When a ROM, constructed based on the POD basis obtained from training data, is applied to new parameter settings, the ROM often lacks of robustness against the change of parameters.
Indeed, the lack of robustness against the change of parameters limits the application of ROMs to design, control, and other real-time operations problems which typically involve parameter changes. In this paper, \textit{we propose to use regression trees to learn the mapping between parameters and the POD bases that span the low-dimensional subspaces onto which full-order models are projected}. Once the mapping has been learned, the tree-based model can automatically select the reduced-order basis and thus adapt ROMs for new parameter settings.

\subsection{Motivating Examples} \label{sec:motivation}
In this paper, five examples are presented to comprehensively investigate the performance of the proposed approach (see Section \ref{sec:numerical}). In this section, only one of the five examples is described in detail so as to demonstrate the needs for adapting ROMs with respect to parameter changes. 

Airborne collisions between aircraft and Unmanned Aerial Vehicle (UAV) has been identified as an emerging threat to aviation safety by the Federal Aviation Administration \citep{FAA2017, FAA2020}. To understand the collision severity under different collision parameters, high-fidelity Finite Element Analysis (FEA) is used to simulate the collision processes. 
In this example, we consider the aircraft nose metal skin deformation process due to UAV collisions at different impact attitudes (i.e., pitch, yaw and roll degree).
UAV's attitudes often cause different aircraft surface skin deformation because the flight attitudes determine the components that firstly hit the aircraft. 
 FEA is performed under 35 different combinations of pitch, yaw and roll degrees and each parameter is chosen from an interval $[-45^{\circ}, 45^{\circ}]$.
The length of each simulation time step is set to 0.02 millisecond, and the impact velocity is fixed to 151$\text{m} \cdot \text{s}^{-1}$ along the impact direction. In the aircraft nose finite element model, structural parts are modeled with 691,221 shell elements and the average size of mesh is 14mm.  In the UAV finite element model, 5,044 solid and 8,900 shell elements are incorporated because some UAV parts like motors and battery cannot be modeled by shell elements.

\begin{figure}[h!]
     \centering
         \includegraphics[width=\textwidth]{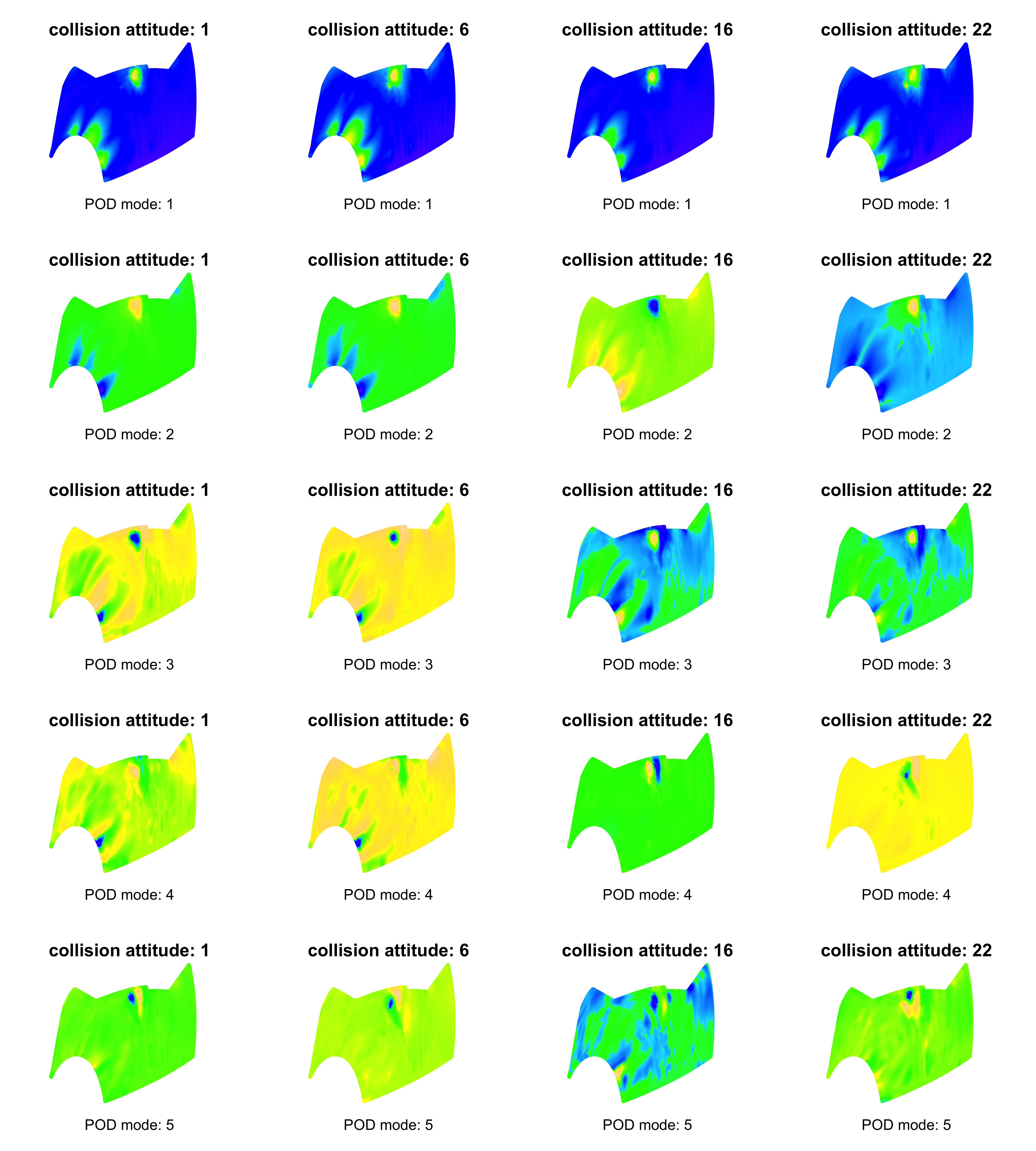}        
     \vspace{-8pt}
     \caption{POD bases for 4 selected collision conditions. Condition 1: pitch ($0^{\circ}$), yaw ($-3^{\circ}$), roll ($33^{\circ}$); Condition 6: pitch ($29^{\circ}$), yaw ($44^{\circ}$), roll ($-13^{\circ}$); Condition 16: pitch ($-10^{\circ}$), yaw ($-13^{\circ}$), roll ($-15^{\circ}$); Condition 22: pitch ($-45^{\circ}$), yaw ($37^{\circ}$), roll ($-12^{\circ}$).}
     \label{fig:POD}
\end{figure}

Using the snapshot data generated from FEA, a ROM can be constructed for each collision condition \citep{Benner2015}. Figure \ref{fig:POD} shows the top 5 POD bases for 4 selected collision conditions. We see that, 

$\bullet$ The POD bases can be different under different collision parameters. Hence, if a ROM is constructed using the POD basis obtained from the training data, such a model becomes less robust and accurate when it is applied to other collision parameters. This is the main motivation behind interpolating POD bases so as to adapt ROMs for different parameter settings \citep{Amsallem2008, Benner2015}. 

$\bullet$ The global POD basis, which is constructed by combining all snapshot data from all parameter settings, may not be reliable for individual parameters. 

$\bullet$ It is also noted that, for certain parameter settings, their corresponding POD bases do appear to be similar. For example, the first and second columns of  Figure \ref{fig:POD} corresponding to collision conditions 1 and 6. 

These observations raise an important question: \textit{is it possible to use statistical learning method to partition the parameter space into a number of sub-regions, and then construct the ``locally-global'' POD basis for each sub-region over the parameter space?} In Section \ref{sec:numerical}, we present five examples to demonstrate the possibility of this idea, including the heat equation, Schrodinger's equation, Thomas Young's double slit experiment, temperature field for Electron Beam Melting (EBM) additive manufacturing processes, and aircraft-UAV collision processes. As we can see in that section, the answer to the question above is not always straightforward. But when it is possible to learn such a mapping from parameters to POD bases, one is able to better adapt ROMs for different parameter settings.  

\subsection{Literature Review and the Proposed Research}

To adapt ROMs for different parameter settings, different approaches have been proposed in the literature, such as the use of global POD,  interpolation of POD basis, machine learning approaches (e.g., Gaussian process), and the construction of a library POD bases. 

The global POD approach constructs one universal POD basis and apply the universal POD basis to construct ROMs under all parameter settings \citep{Benner2015, Mak2019, Qian2020}. Hence, for the global POD to work well, it is necessary to generate data from a relatively large number of parameter settings from the parameter space. Generating such a large dataset from multiple parameter settings can be expensive (for example, consider solving a finite element problem or a fluid dynamics problem numerically). In addition, the global POD is no longer optimal for individual parameters and may become inaccurate for certain parameter settings as illustrated by Figure \ref{fig:POD}. However, it is also worth noting that, the use of global POD works well in one of our five examples in Section \ref{sec:numerical} when the mapping between parameters and POD bases can hardly be learned using other alternative approaches. 

The interpolation approach directly interpolates the POD basis for a new parameter setting \citep{Lieu2007, Amsallem2008, Benner2015, Friderikos2020}. One popular approach is to interpolate the POD basis via Grassmann manifolds \citep{Amsallem2008}. Based on this approach, the sub-space spanned by a POD basis is viewed as a point in the Grassmann manifold \citep{Boothby2002}. Then, these sub-spaces are mapped to a flat tangent space, and the interpolation is performed on the flat tangent space. After that, the interpolated value on the tangent space is mapped back to the Grassmann manifold which yields the interpolated POD basis. This approach effectively maintains the orthogonality property of the reduced-order basis. The stability of this approach has been investigated by \cite{Friderikos2020}. The computational advantage of this approach also enables near real-time adaptation of ROM. 

Constructing a library of physics-based ROM is another approach for adapting ROMs. For example, \cite{Amsallem2009} proposed on-demand CFD-based aeroelastic predictions which rely on the pre-computation of a database of reduced-order bases and models for discrete flight parameters. For adapting digital twins models under changing operating conditions, \citep{Kapteyn2020} used classification trees to classify the sensor data and then select the appropriate physics-based reduced models from the model library. Building a library of ROM itself could be time-consuming and may not always be feasible for certain applications. In recent years, we have also seen some work using machine learning approaches to predict the subspace onto which the full-order model is projected. For example, \cite{https://doi.org/10.48550/arxiv.2107.04668} proposed a Bayesian nonparametric Gaussian Process Subspace regression (GPS) model for subspace prediction. With multivariate Gaussian distributions on the Euclidean space, the method hinges on the induced joint probability model on the Grassmann manifold which enables the prediction of POD basis. \cite{Giovanis_2020} extended the interpolation method by utilizing the unsupervised clustering approach to cluster data into different clusters within which solutions are sufficiently similarly such that they can be interpolated on the Grassmannian. 


The idea presented in this paper \textit{resides between directly interpolating POD basis and constructing global POD basis}. The paper proposes a binary regression tree to optimally partition the parameter space into a number of sub-regions, and construct the ``locally-global'' POD bases for each sub-region. 
Recall that, a sub-space spanned by a POD basis can be viewed as a point in the Grassmann manifold, which enables us to leverage the Riemannian distance between two sub-spaces (i.e., points) \citep{Friderikos2020}. 
In our regression tree, the points contained in a tree node are precisely the sub-spaces spanned by POD bases. Given any tree node split parameter and split value, it is possible to partition the sub-spaces in a tree node into the left and right daughter nodes. For each daughter node, we compute the global POD only using the data within that daughter node (we call it the ``locally-global'' POD) and the corresponding sub-space spanned by the locally-global POD basis. This allows us to compute the sum of the Riemannian distances, within a daughter node, between the individual sub-spaces contained in the node and the sub-space spanned by the locally-global POD basis on that node. Hence, the optimal tree node splitting parameter and split value can be found by minimizing the sum of the Riemannian distances on the two daughter nodes. Repeating this tree node splitting procedure gives rise to the well-known greedy algorithm which has been widely used to grow binary regression trees \citep{Hastie2016}. 

Note that, such a ``divide-and-conquer'' idea above can be traced back to \cite{Du2003}, which utilized the Voronoi tessellations (CVT) as a clustering technique to systematically extract best POD basis. \cite{Eftang2010} also proposed a hierarchical splitting of the parameter domains based on proximity to judiciously chosen parameter anchor points within each subdomain. This idea, in a nutshell, is similar to the use of regression trees to partition the feature space, and construct POD basis for each sub feature space. 
In contrast, when growing the regression tree on Grassmann manifold, the partition of feature space and construction of POD bases for sub feature spaces become a single integrated step driven by minimizing a pre-determined loss function (in this paper, the total Riemannian distances between the subspaces spanned by POD bases on tree leaves). The simplicity and excellent interpretability of regression trees enable the transparency, fast computation and easy implementation of the proposed method. To our best knowledge, regression trees on Grassmann manifold for adapting ROM has not been developed in the literature. 

In Section \ref{sec:tree}, we present the technical details of how a regression tree for adapting ROM can be grown. In Section \ref{sec:numerical}, numerical examples are presented to demonstrate the potential advantages of the proposed approach and compare the performance of different approaches. Section \ref{sec:conclusion} concludes the paper. 

\section{Regression Tree for Adapting ROM} \label{sec:tree}

\subsection{The Problem} \label{sec:problem}

Parametric model reduction is used to generate computationally faster ROMs that approximate the original systems. This usually starts with generating data from the full-order system by numerically solving the governing PDE at a small set of parameter settings, $\Lambda = \{\bm{\lambda}_1, \bm{\lambda}_2, \cdots, \bm{\lambda}_N\}$ for $N \in \mathbb{N}^+$. Here, $\bm{\lambda}_i = (\lambda^{(1)}_i, \lambda^{(2)}_i, \cdots, \lambda^{(d)}_i) \in \mathbb{P}$ where $\mathbb{P} \in \mathbb{R}^d$ is the $d-$dimensional parameter space. For instance, in the example described in Section \ref{sec:motivation}, FEA is used to simulate the aircraft nose surface deformation at $N=35$ collision conditions, and $\bm{\lambda}_i$ contains the pitch, yaw and roll degrees under each collision condition $i$. 

Given the parameter $\bm{\lambda}$, the solution of the full-order system is often a space-time field:
\begin{equation}
(t,\bm{s}) \in [0,T] \times \mathbb{S} \mapsto x(t, \bm{s}; \bm{\lambda})
\end{equation}
where $\mathbb{S}$ is the spatial domain, $T>0$ and $x(t, \bm{s}; \bm{\lambda})$ is the solution of the full-order system given the parameter  $\bm{\lambda}$. For example, $x(t, \bm{s}; \bm{\lambda})$ can be the surface deformation at location $\bm{s}$ and time $t$ under the collision condition $\bm{\lambda}$.  
Let $\bm{x}(t; \bm{\lambda}) = (x(t, \bm{s}_1; \bm{\lambda}), x(t, \bm{s}_2; \bm{\lambda}), \cdots, x(t, \bm{s}_n; \bm{\lambda}))^T$ be the solutions $x(t, \bm{s}; \bm{\lambda})$ at discrete locations $\bm{s}_1,\bm{s}_2,\cdots,\bm{s}_n$ and at time $t$ given the parameter $\bm{\lambda}$, then, 
\begin{equation}
\bm{D}(\bm{\lambda}) = [\bm{x}(t_1; \bm{\lambda}), \bm{x}(t_2; \bm{\lambda}), \cdots, \bm{x}(t_{n_T}; \bm{\lambda})]
\end{equation}
is called the \textit{snapshot matrix} (i.e., data) produced by numerically solving the full-order physics at a parameter setting $\bm{\lambda}$. 

In many engineering and scientific applications, the dimension $n$ of the vector $\bm{x}(t; \bm{\lambda})$ can be extremely high, and it is computationally impossible to solve the full-order physics model for all potential parameter settings. Hence, the projection-based model reduction seeks for a $r$-dimensional vector ($r \ll  n$) such that $\bm{x}(t; \bm{\lambda}) \approx \bm{\Phi}_{\bm{\lambda}} \bm{x}_r(t; \bm{\lambda})$, where $\bm{\Phi}_{\bm{\lambda}} =(\bm{\phi}_1, \bm{\phi}_2,\cdots,\bm{\phi}_r) \in \mathbb{R}^{n\times r}$ is the matrix basis with orthogonal column vectors. Here, the basis matrix $\bm{\Phi}$ belongs to the compact Stiefel manifold $\mathcal{ST}(r,n)$, which is a set of all $n\times r$ matrices of rank $r$. 

The optimal matrix basis $\bm{\Phi}_{\bm{\lambda}}$ for parameter $\bm{\lambda}$ is found by 
\begin{equation} \label{eq:min}
\bm{\Phi}_{\bm{\lambda}} =  \text{argmin}_{\bm{\Phi}} || \bm{D}(\bm{\lambda}) - \bm{\Phi}\bm{\Phi}^T\bm{D}(\bm{\lambda}) ||_F^2
\end{equation}
where $||\cdot||_F$ is the Frobenius norm on the vector space of matrices. The solution of this minimization problem is well-known according to the Eckart-Young Theorem and is obtained by the Singular Value Decomposition (SVD) of the snapshot matrix $\bm{D}(\bm{\lambda})$ \citep{Benner2015, Mak2019, Qian2020}:
\begin{equation} \label{eq:SVD}
\bm{D}(\bm{\lambda}) = \bm{U}\bm{\Sigma}\bm{V}^T, \quad\quad  \bm{U} \equiv [\bm{u}_1, \bm{u}_2, \cdots, \bm{u}_{n_T}]
\end{equation}
and $\bm{\Phi}_{\bm{\lambda}} = [\bm{u}_1, \bm{u}_2, \cdots, \bm{u}_r]$.


Hence, it is clearly seen that the POD basis $\bm{\Phi}_{\bm{\lambda}}$ is only locally optimal for the parameter setting $\bm{\lambda}$ and may lack robustness over parameter changes, i.e., it could be unreliable to apply the POD basis, obtained from a limited set of parameters (training set), to new parameter settings (testing set) in the parameter space, leading to the following problem statement.

\vspace{4pt}
\textit{Problem Statement}. Given $\{(\bm{\lambda}_i, \bm{\Phi}_{\bm{\lambda}_i})\}_{i=1}^{N}$ where $\bm{\Phi}_{\bm{\lambda}_i}$ is the POD basis constructed for parameter setting $\bm{\lambda}_i \in \Lambda$, the problem is concerned with learning a mapping $\mathcal{T}: \mathbb{P} \mapsto \mathcal{ST}(r,n)$ from the parameter space $\mathbb{P} \in \mathbb{R}^d$ to the compact Stiefel manifold. If such a mapping $\mathcal{T}$ can be constructed, it is possible to predict the POD basis $\bm{\Phi}_{\bm{\lambda}^*}$ for a new parameter setting $\bm{\lambda}^* \notin \Lambda$ (without numerically solving the computationally expensive full-order physics model at $\bm{\lambda}^*$). 

\begin{figure}[h!]
     \centering
         \includegraphics[width=\textwidth]{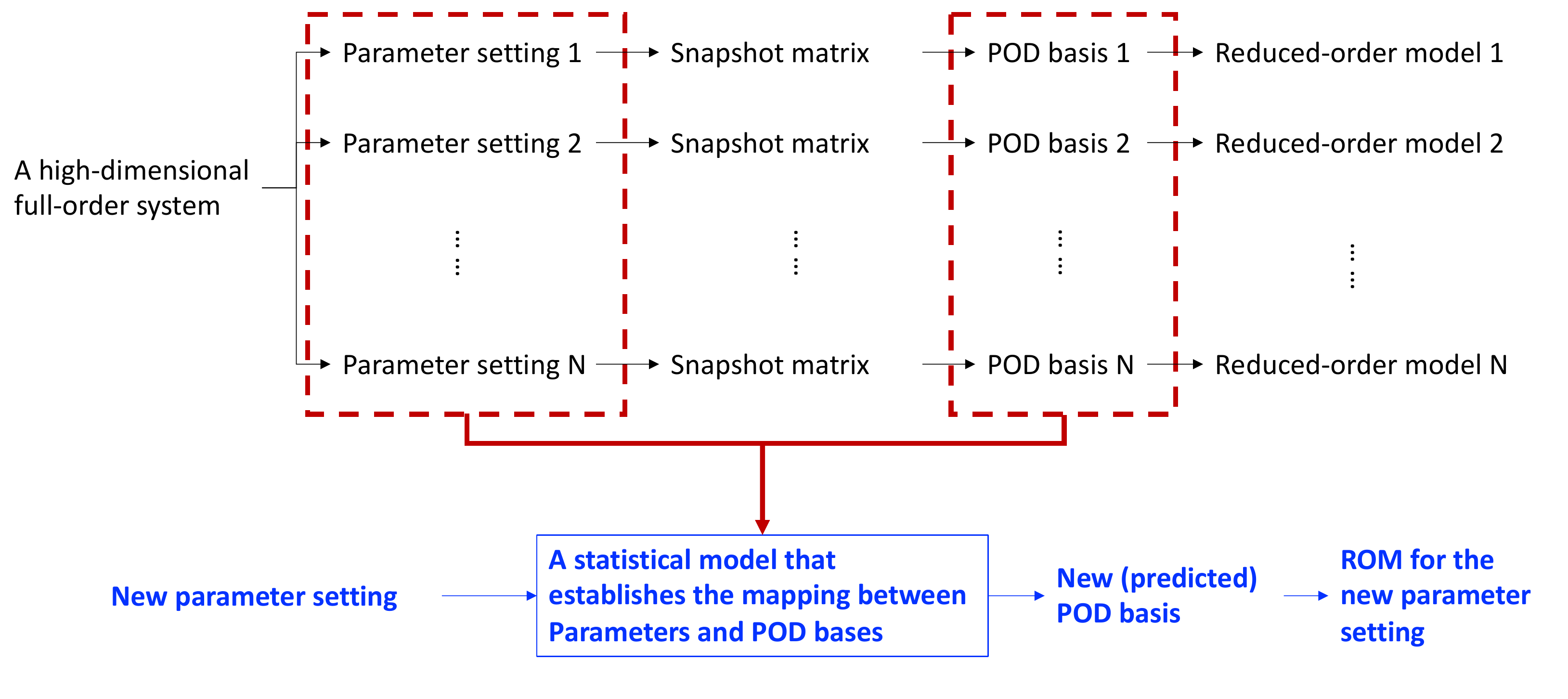}        
     \vspace{-8pt}
     \caption{An illustration of the problem: learning the mapping between parameters and POD bases such that it is possible to predict the POD bases for new parameter settings.}
     \label{fig:problem}
\end{figure}

The problem statement is illustrated in Figure \ref{fig:problem}. 

\subsection{Regression Tree on Grassmann Manifold} \label{sec:problem}

Binary regression trees are used to establish the mapping between $\bm{\lambda}$ and $\bm{\Phi}_{\bm{\lambda}}$. Let $\mathbb{P}$ be the parameter space of $\bm{\lambda}$. Suppose a regression tree divides the parameter space $\mathbb{P}$ into $M$ regions, $R_1, R_2, \cdots, R_M$, the predicted basis $\hat{\bm{\Phi}}^*$ for a new parameter $\bm{\lambda}^* \notin \Lambda$ is given by
\begin{equation} \label{eq:tree}
\hat{\bm{\Phi}}^* = f(\bm{\lambda}^*) = \sum_{m=1}^{M} \hat{\bm{\Phi}}_m I_{\{\bm{\lambda}^* \in R_m\}}
\end{equation}
where $\hat{\bm{\Phi}}_m$ is the locally-global POD basis constructed using data only from region $m$. In particular, let 
\begin{equation} 
\bm{D}_m = [\bm{D}(\bm{\lambda}_1)I_{\{\bm{\lambda}_1 \in R_m\}}, \bm{D}(\bm{\lambda}_2)I_{\{\bm{\lambda}_2 \in R_m\}}, \cdots, \bm{D}(\bm{\lambda}_N)I_{\{\bm{\lambda}_N \in R_m\}}]
\end{equation}
be the snapshot matrix from region $m$, and $\hat{\bm{\Phi}}_m$ is obtained from the SVD of $\bm{D}_m$; see (\ref{eq:SVD}).

Like how regression and classification trees are typically grown, we proceed with the greedy algorithm. At each tree node, we find the optimal splitting variable $j$ and split point $l$ that define the left and right daughter nodes:
\begin{equation} 
R_L(j,l) = \left\{ \bm{\lambda}| \lambda^{(j)} \leq l  \right\},  \quad\quad\quad
R_R(j,l) = \left\{ \bm{\lambda}| \lambda^{(j)} > l  \right\}
\end{equation}
where $\lambda^{(j)}$ is the $j$-th parameter in $\bm{\lambda}$. 

Using the data on the left and right daughter nodes, we respectively compute the locally-global POD bases $\hat{\bm{\Phi}}_L$ and $\hat{\bm{\Phi}}_R$, and the optimal splitting variable $j$ and split point $l$  are found by minimizing
\begin{equation}  \label{eq:min}
\text{min}_{j,l} \left\{ \sum_{\lambda^{(j)}_i \in R_L(j,l)} \delta(\bm{\Phi}_{\bm{\lambda}_i} , \hat{\bm{\Phi}}_L) + \sum_{\lambda^{(j)}_i \in R_R(j,l)} \delta(\bm{\Phi}_{\bm{\lambda}_i} , \hat{\bm{\Phi}}_R)\right\}
\end{equation}
where $\lambda^{(j)}_i$ is the $j$-th element in $\bm{\lambda}_i$, and $\delta(\bm{\Phi},\bm{\Phi}')$ is some distance measure between the two POD bases $\bm{\Phi}$ and $\bm{\Phi}'$. 

\textit{Riemannian distance}. The choice of the distance measure $\delta(\cdot,\cdot)$ becomes critical. Note that, the distance between any two POD bases cannot be measured in a linear space (for example, we cannot directly compute the Frobenius distance between the two basis matrices). This is because each POD basis matrix $\bm{\Phi}_{\bm{\lambda}_i} \in \mathbb{R}^{n\times r}$ spans a $r$-dimensional vector subspace $\mathcal{S}_{\bm{\Phi}}$ onto which the original state vector $\bm{x}(t; \bm{\lambda})$ is projected. Hence, one natural way to define the distance measure $\delta(\cdot,\cdot)$ is to consider the \textit{Riemannian distance} between the two vector subspaces, $\mathcal{S}_{\bm{\Phi}}$ and $\mathcal{S}_{\bm{\Phi}'}$, respectively spanned by POD bases $\bm{\Phi}$ and $\bm{\Phi}'$. 

Note that, the matrix $\bm{\Phi}$ belongs to the compact Stiefel manifold $\mathcal{ST}(r,n)$, which is a set of all $n\times r$ matrices of rank $r$, and the $r$-dimensional vector subspace $\mathcal{S}_{\bm{\Phi}}$ can be viewed as a point that belongs to the Grassmann manifold $\mathcal{G}(r,n)$, 
\begin{equation} 
\mathcal{G}(r,n) = \{\mathcal{S}_{\bm{\Phi}} \subset \mathbb{R}^n, \text{dim}(\mathcal{S}_{\bm{\Phi}})=r \}
\end{equation}
which is a collection of all $r$-dimensional subspaces of $\mathbb{R}^n$. This enables us to properly define the Riemannian distance between the two subspaces $\mathcal{S}_{\bm{\Phi}}$ and $\mathcal{S}_{\bm{\Phi}'}$ spanned by $\bm{\Phi}$ and $\bm{\Phi}'$, and use such a distance as $\delta(\cdot,\cdot)$:
\begin{equation}  \label{eq:delta_1}
\delta^{\text{Riemannian}}(\bm{\Phi},\bm{\Phi}') \equiv  \left( \sum_{i=1}^{r} \text{arccos}(\sigma_i^2) \right)^{\frac{1}{2}} \equiv \left( \sum_{i=1}^{r} \theta_i^2 \right)^{\frac{1}{2}}
\end{equation}
where $\theta_i = \text{arccos}(\sigma_{r-i+1})$ is the \textit{Jordan's principal angles} between $\bm{\Phi}$ and $\bm{\Phi}'$, and $0 \leq \sigma_r  \leq \cdots \leq \sigma_1$ are the singular values of  $\bm{\Phi}^T \bm{\Phi}'$ \citep{Friderikos2020}. 

\begin{figure}[h!]
     \centering
         \includegraphics[width=0.8\textwidth]{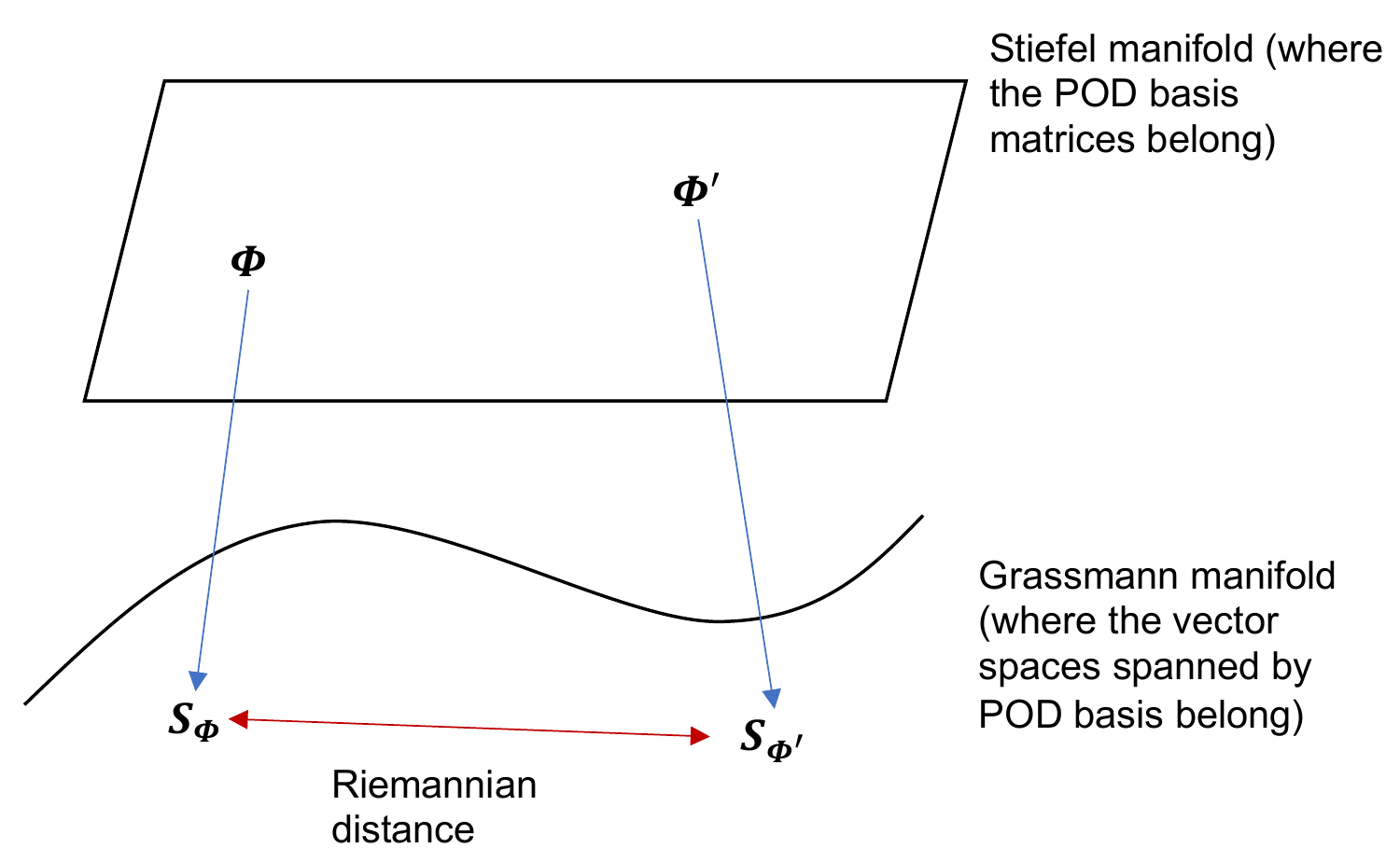}        
     \vspace{-8pt}
     \caption{The Riemannian distance between the two vector spaces $\mathcal{S}_{\bm{\Phi}}$ and $\mathcal{S}_{\bm{\Phi}'}$ (i.e., two points in the Grassmann manifold) is used as a distance measure between $\bm{\Phi}$ and $\bm{\Phi}'$ (i.e., two points in the Stiefel manifold)}
     \label{fig:Grassmann}
\end{figure}

The idea is sketched in Figure \ref{fig:Grassmann}. Given any POD basis matrices $\bm{\Phi}$ and $\bm{\Phi}'$ (i.e., two points in the Stiefel manifold), we use the Riemannian distance between the two vector spaces $\mathcal{S}_{\bm{\Phi}}$ and $\mathcal{S}_{\bm{\Phi}'}$ (i.e., two points in the Grassmann manifold) as a proper distance measure between $\bm{\Phi}$ and $\bm{\Phi}'$. Note that, because a POD basis matrix spans a vector space onto which the full-order physics is projected, it is not meaningful to directly compute the difference between two POD basis matrices. 

\textit{Computational considerations}. The discussions above explain how a regression tree can be grown on the Grassmann manifold, but it is noted that growing such a tree can be computationally intensive. This is because solving the optimization problem (\ref{eq:min}) requires repeatedly performing SVD for candidate splitting variables and split points. Fortunately, if the snapshot data do have a low-rank structure, then, the \textit{randomized} SVD (rSVD) can be adopted at a fraction of the cost of conventional SVD \citep{brunton2022data}.

In particular, let $\bm{D}$ be the snapshot data matrix at a tree node, we randomly sample the column space of $\bm{D}$ using a random project matrix $\bm{P}$ with $r'$ columns
\begin{equation} 
\bm{D}' = \bm{D}\bm{P}.
\end{equation}
Here, $r'$ is much smaller than the number of columns in $\bm{D}$, and $\bm{D}'$ approximates the column space of $\bm{D}$. 
Then, the low-rank QR decomposition of $\bm{D}'=\bm{Q}\bm{R}$ provides an orthonormal basis for $\bm{D}$. It is possible to project $\bm{D}$ to the low-dimensional space spanned by $\bm{Q}$, $\tilde{\bm{D}}=\bm{Q}^*\bm{D}$ (where $\cdot^*$ is the complex conjugate transpose), and perform the SVD on a much smaller matrix $\tilde{\bm{D}}$
\begin{equation} 
\tilde{\bm{D}} = \bm{U}_{\tilde{\bm{D}}} \bm{\Sigma}_{\tilde{\bm{D}}} \bm{V}_{\tilde{\bm{D}}}^*
\end{equation}
Because $\tilde{\bm{D}}$ is obtained by multiplying a matrix $\bm{Q}^*$ to $\bm{D}$ from the left, it is well-known that $\bm{\Sigma}_{\tilde{\bm{D}}}=\bm{\Sigma}_{\bm{D}}$, $\bm{V}_{\tilde{\bm{D}}}=\bm{V}_{\bm{D}}$, and the left singular vector $\bm{U}_{\bm{D}}$ of  $\bm{D}$ is given by
\begin{equation} 
\bm{U}_{\bm{D}} =\bm{Q} \bm{U}_{\tilde{\bm{D}}}.
\end{equation}

\section{Numerical Examples} \label{sec:numerical}

In this section, we present a comprehensive numerical investigation on the performance  of the proposed approach. Five examples are considered, including the heat equation, Schrodinger's equation, Thomas Young's double slit experiment, temperature field for Electron Beam Melting (EBM) additive manufacturing, and aircraft nose deformation due to UAV collisions. Comparison studies are also performed to compare the performance of different approaches.

\subsection{Example I: heat equation}
\begin{figure}[h!] 
     \centering
         \includegraphics[width=0.9\textwidth]{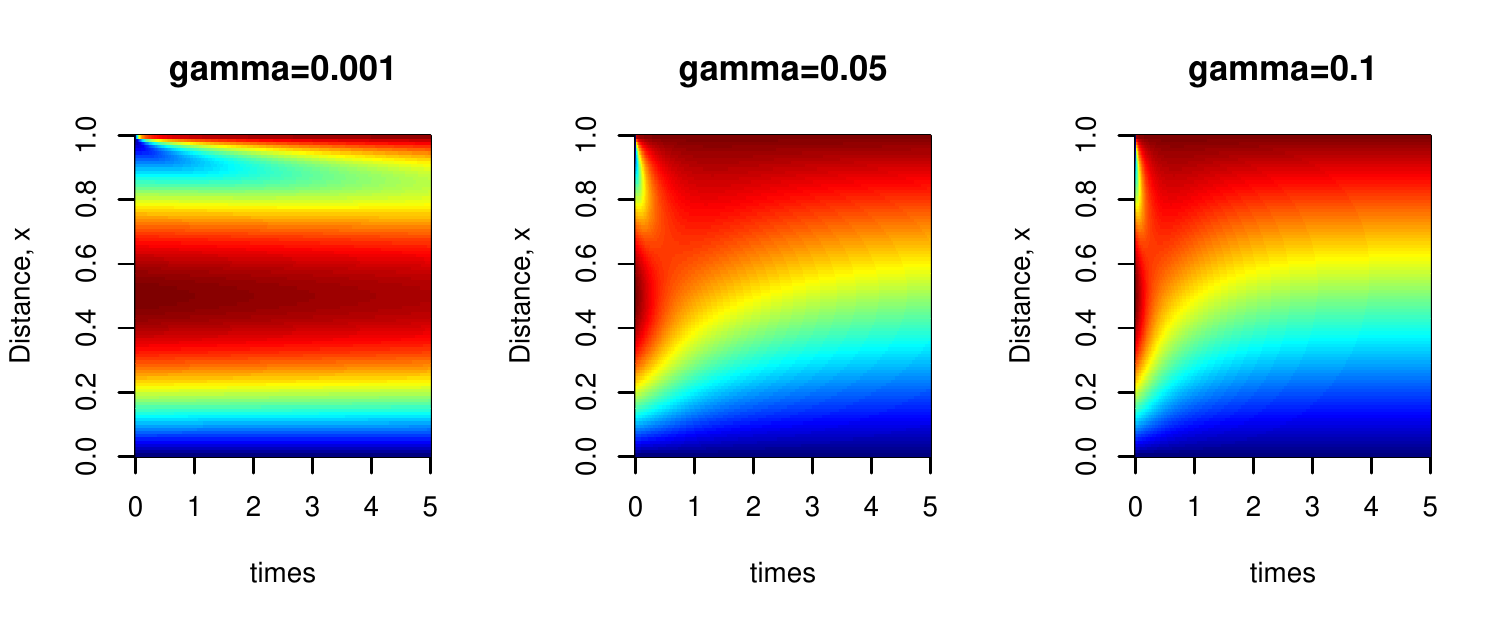}        
     \vspace{-8pt}
     \caption{Solutions of a 1D heat equation for different thermal diffusivity parameters.}
     \label{fig:heat}
\end{figure}

In example I, we start with a simple 1D heat equation
\begin{equation} \label{eq:heat}
\xi_t = \gamma^2 \xi_{xx}, \quad\quad\quad  \xi(0,x)=\sin(\pi x), x\in[0,1],  t \geq 0
\end{equation}
where $\xi(t,x)$ is the temperature distribution in time and on a unit interval $[0,1]$, and $\gamma$ is a positive parameter representing the thermal diffusivity of the medium. We solve the heat equation (\ref{eq:heat}) for $t\in[0,5]$ at 100 different values of $\gamma \in \Omega_{\gamma}=\{ 0.001, 0.002, \cdots, 0.099, 0.1\}$. For any given $\gamma$, the obtained snapshot data are denoted by $\bm{D}(\gamma)$. 
Figure \ref{fig:heat} shows the temperature distribution in time and space for $\gamma=0.001$, $\gamma=0.05$ and $\gamma=0.1$. We see that the solution of the heat equation depends on the thermal diffusivity parameter $\gamma$. 

In our experiment, we train the tree using the data from 21 values of $\gamma$ from the training set 
\begin{equation} 
\begin{split}
\Omega_{\gamma}^{\text{train}} = \{ & 0.001, 0.006, 0.011, 0.016, 0.021, 0.026, 0.031, 0.036, 0.041, \\ 
& 0.046, 0.051, 0.056, 0.061, 0.066, 0.071, 0.076, 0.081, 0.086, 0.091, 0.096, 0.1\}, 
\end{split}
\end{equation}
and the tree is then used to predict the POD bases for the remaining 79 parameter settings for $\gamma$ from the testing set $\Omega_{\gamma}^{\text{test}} = \Omega_{\gamma}\setminus\Omega_{\gamma}^{\text{train}}$. Finally, the error $e_{\gamma}$, $\gamma \in \Omega_{\gamma}$, is measured by the Frobenius norm:
\begin{equation} 
e_{\gamma} = || \bm{D}(\gamma) - \hat{\bm{\Phi}}_{\gamma} \hat{\bm{\Phi}}_{\gamma}^T\bm{D}(\gamma) ||_F^2, \quad\quad \text{for } \gamma \in \Omega_{\gamma}^{\text{test}}
\end{equation}
where $\hat{\bm{\Phi}}_{\gamma}$ is the predicted POD basis for $\gamma \in \Omega_{\gamma}^{\text{test}}$. 

\begin{figure}
     \centering
     \begin{subfigure}[b]{0.48\textwidth}
         \centering
         \includegraphics[width=\textwidth]{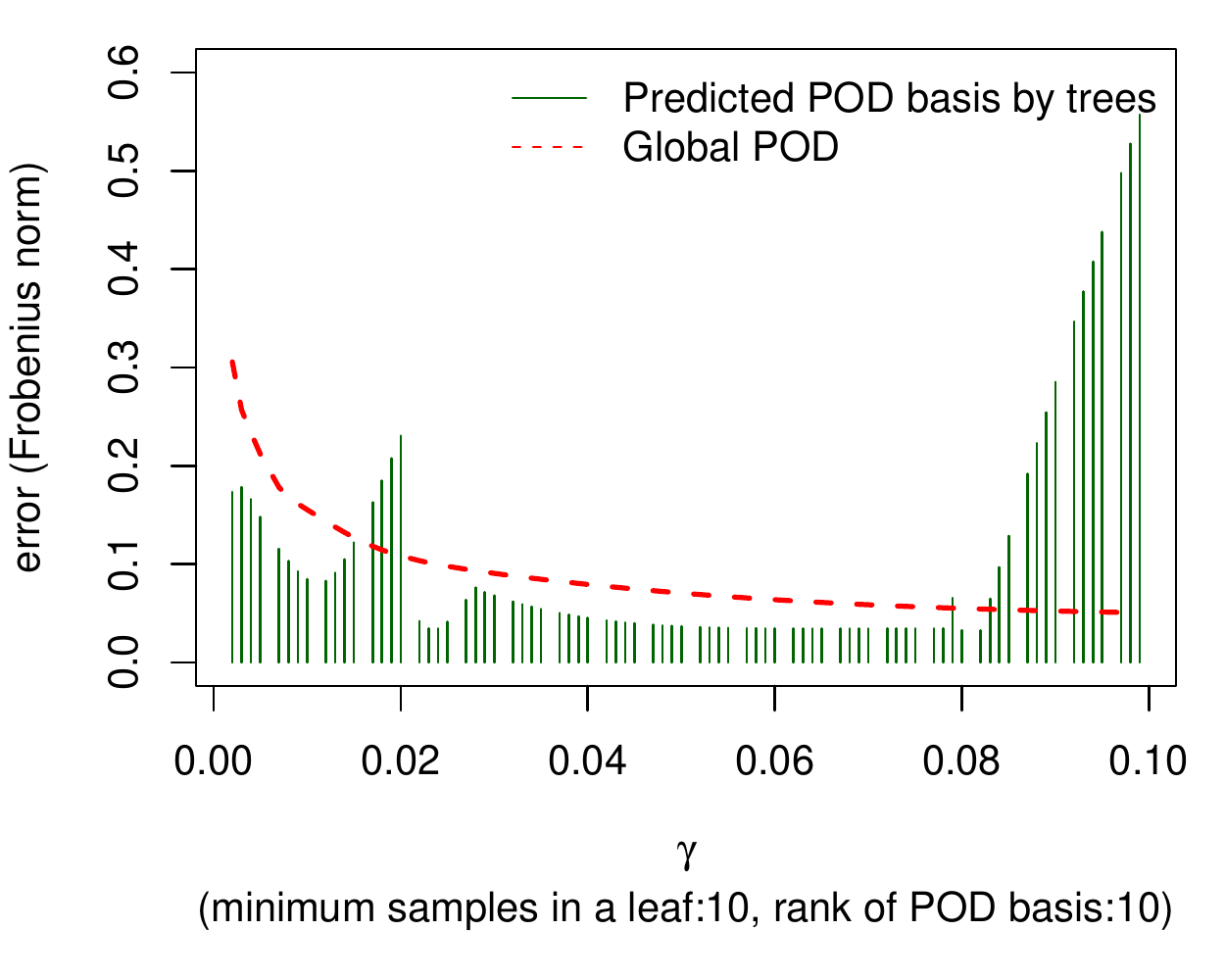}
         \caption{}
     \end{subfigure}
       \hfill
       \begin{subfigure}[b]{0.48\textwidth}
         \centering
         \includegraphics[width=\textwidth]{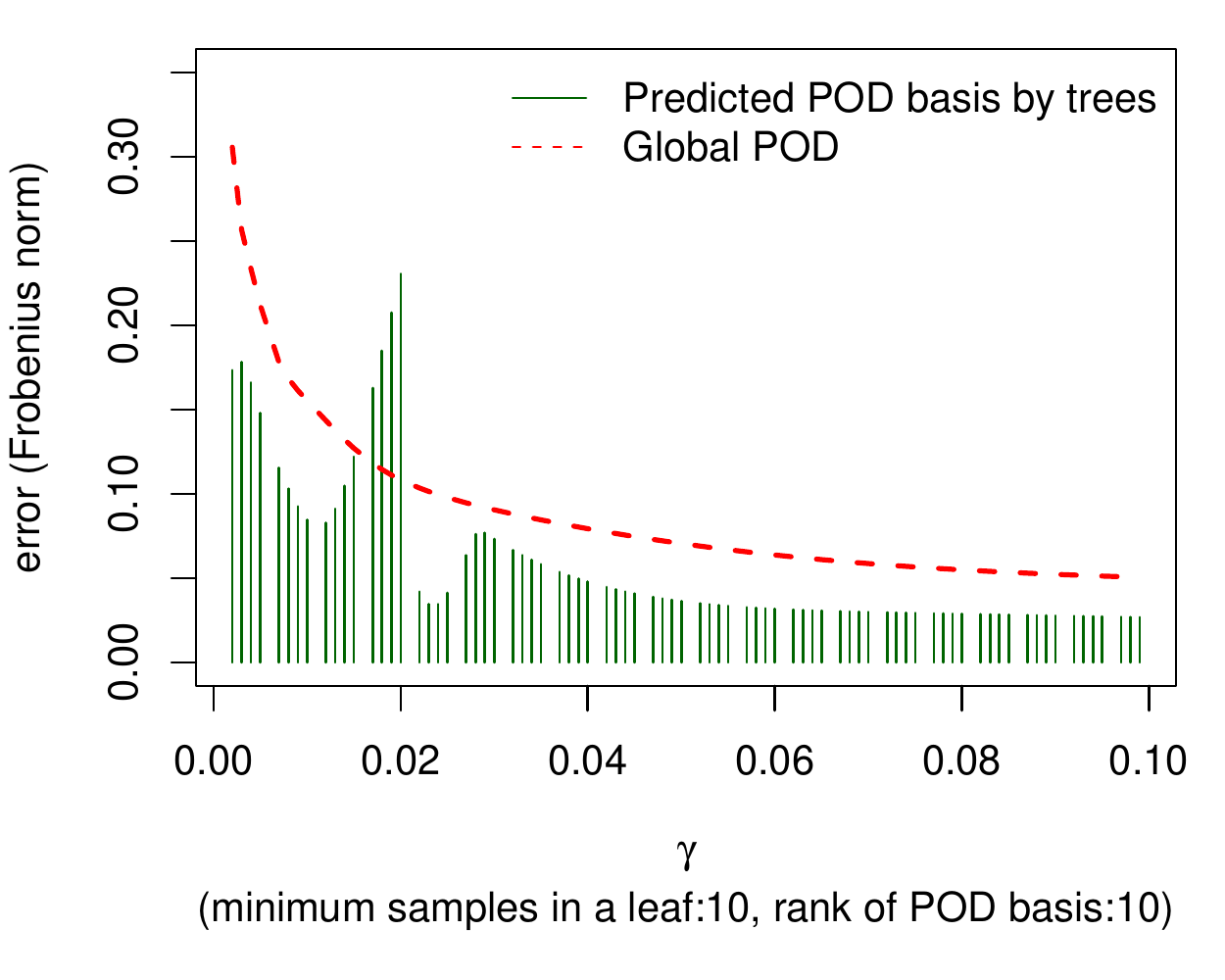}
         \caption{}
     \end{subfigure}
        \caption{The error, $e_{\lambda}$ for $\lambda \in \Omega_{\lambda}$, based on both the predicted POD basis by the proposed tree and the global POD basis.}
        \label{fig:heat_result}
\end{figure}

Figure \ref{fig:heat_result} shows the error, $e_{\gamma}$ for all $\gamma \in \Omega_{\gamma}$, based on both the predicted POD bases by the proposed regression tree and the global POD basis. In particular, we consider different minimum number of samples within a tree leaf (which controls the tree depth and affects the tree performance) and set the rank of the POD basis to 10. We see that, the tree depth affects the performance of the proposed tree (as in the case of conventional regression trees). When the minimum number of samples within a tree leaf is set to 15 (Figure \ref{fig:heat_result}(b)), the prediction error of the proposed tree is lower than that using the global POD for 75 out of the 79 testing cases. When the tree is over complicated (i.e., when we reduce the the minimum number of samples within a tree leaf from 15 to 10), the accuracy of the proposed tree-based method deteriorates due to \textit{over-fitting} the data (Figure \ref{fig:heat_result}(a)).
 
We further compare the proposed method to the existing interpolation method for POD basis via Grassmann manifolds \citep{Amsallem2008}. This approach requires one to select a reference point in the Grassmann manifold (i.e., a subspace spanned by the POD basis constructed at one particular parameter setting from the training dataset), and performs the interpolation on the flat tangent space associated with the selected reference point. In this example, because there are 21 values of $\lambda$ in the training dataset, each of these 21 values can be used to establish the reference point. Hence, we perform the POD basis interpolation for the 79 values of $\lambda$ in the testing dataset for 21 times, and each time a particular value of $\lambda$ in the training dataset is used to establish the reference point. 

\begin{figure}
     \centering
     \begin{subfigure}[b]{0.48\textwidth}
         \centering
         \includegraphics[width=\textwidth]{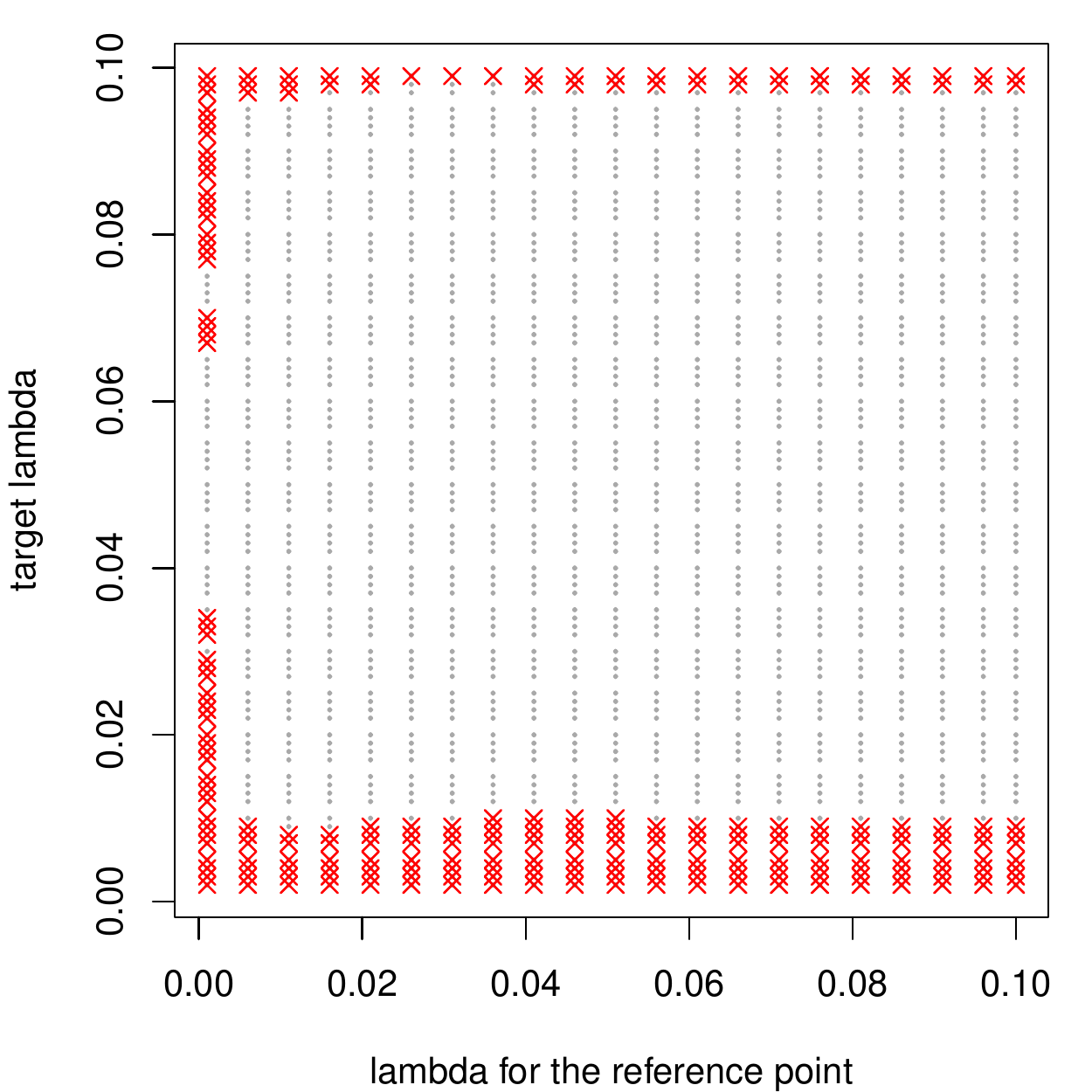}
         \caption{}
         \label{fig:three sin x}
     \end{subfigure}
      \hfill
       \begin{subfigure}[b]{0.48\textwidth}
         \centering
         \includegraphics[width=\textwidth]{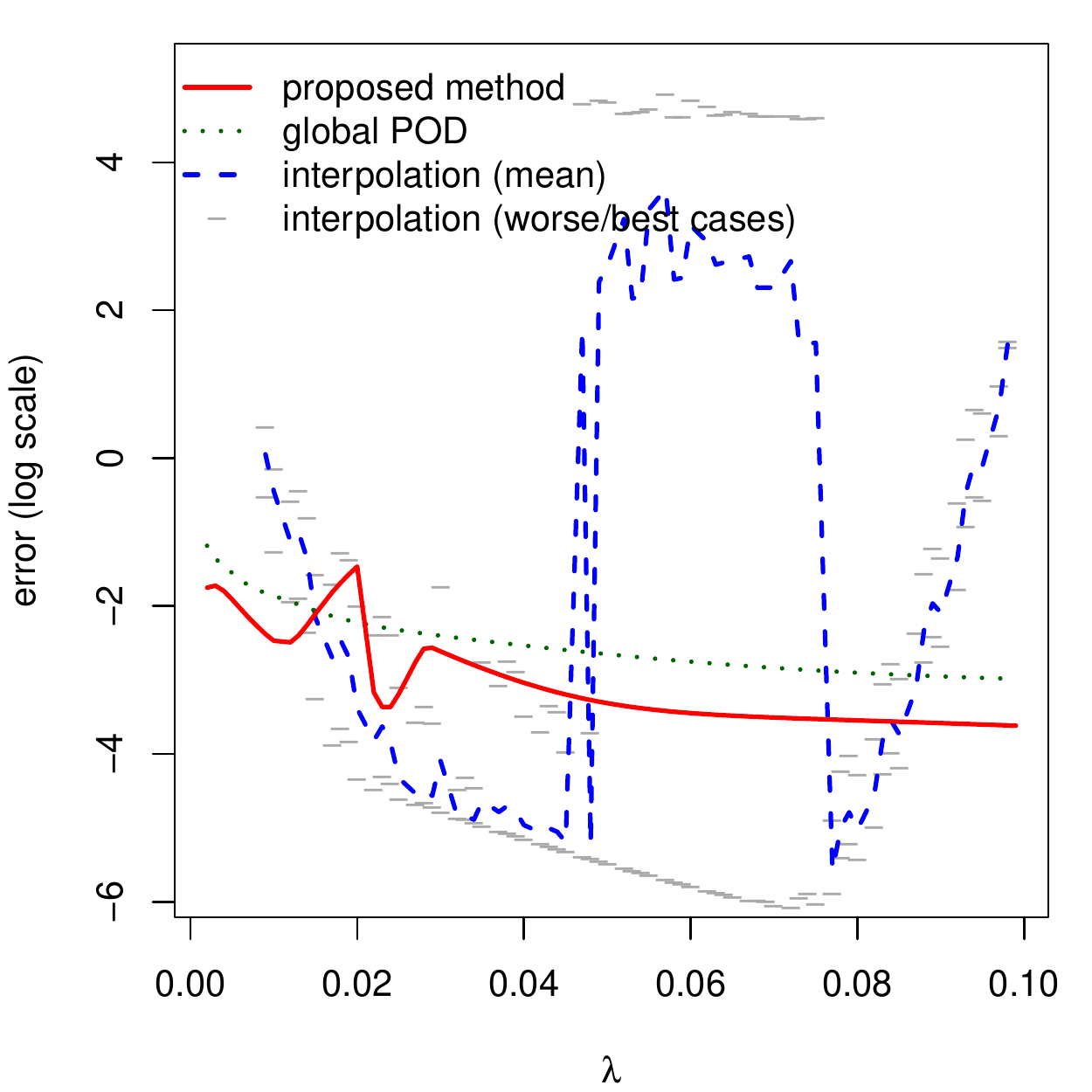}
         \caption{}
         \label{fig:three sin x}
     \end{subfigure}
        \caption{(a) Stability of the POD basis interpolation under the 79 parameter values of $\lambda$ for 21 different choices of the reference point (unstable interpolations are indicted by red crosses); (b) Comparison between the proposed tree-based method, global POD and the interpolation method}
        \label{fig:heat_comparison_all}
\end{figure}

Figure \ref{fig:heat_comparison_all}(a) firstly shows the stability of interpolation based on the stability conditions given in \cite{Friderikos2020}. It is immediately seen that the interpolation is unstable for values of $\lambda$ which lie on the boundaries of the range of $\lambda$ regardless of the choices of reference points (unstable interpolations are indicated by red ``$\times$''). 
After removing the unstable interpolation results, Figure \ref{fig:heat_comparison_all}(b) compares the prediction errors (in log scale) for different values of $\lambda$ between the proposed tree-based method, global POD and the interpolation method. In particular, for the interpolation method, because 21 possible reference points can be chosen for each interpolation, we present the averaged interpolation error (from the 21 choices of reference points), the best (when the most appropriate reference point is chosen), and the worst (when the reference point is poorly chosen) errors. Some useful insights can be obtained from Figure \ref{fig:heat_comparison_all} on how one may choose different methods for predicting POD basis:
(i) The proposed tree-based method outperforms for values of $\lambda$ which lie on the two ends of the range of $\lambda$ (i.e., $\lambda \leq 0.013$ and $\lambda \geq 0.087$ as shown in Figure \ref{fig:heat_comparison_all}(b)), even if the most appropriate choice of the reference point is made for the interpolation method. (ii) When the interpolation method is used, the choice of the reference point is critical. As seen in Figure \ref{fig:heat_comparison_all}(b), for the range $0.049\leq \lambda \leq 0.075$, poor choices of the reference point may lead to extremely unreliable interpolations, while appropriate choices of the reference point yield  accurate interpolations.

\subsection{Example II: Schr\"{o}dinger's equation}
In Example II, we consider the non-linear Schr\"{o}dinger equation that models the instance solitons in optical fibre pulse propagation:
\begin{equation} \label{eq:schrodinger}
\xi_t = i \xi_{xx} + i |\xi|^2\xi
\end{equation}
which has a closed-form solution for a single-soliton:
\begin{equation} \label{eq:schrodinger_s}
\xi(t,x) = \sqrt{2\alpha\gamma^{-1}} \exp(i(0.5vx-t(1/4v^2-\alpha)))\text{sech}(\sqrt{\alpha}(x-vt))
\end{equation}
where $v$ is the speed at which the soliton travels, and $\alpha$ is the parameter that determines the soliton's amplitude. The example presented in \cite{Soetaert2012} considered a fast moving and a slow moving solitons. The first one is initially located at $x=0$ and travels at a speed of $v_1=1$, while the second is initially located at $x=25$ and travels at a much slower speed of $v_2=0.1$. 

\begin{figure}[h!] 
     \centering
         \includegraphics[width=0.9\textwidth]{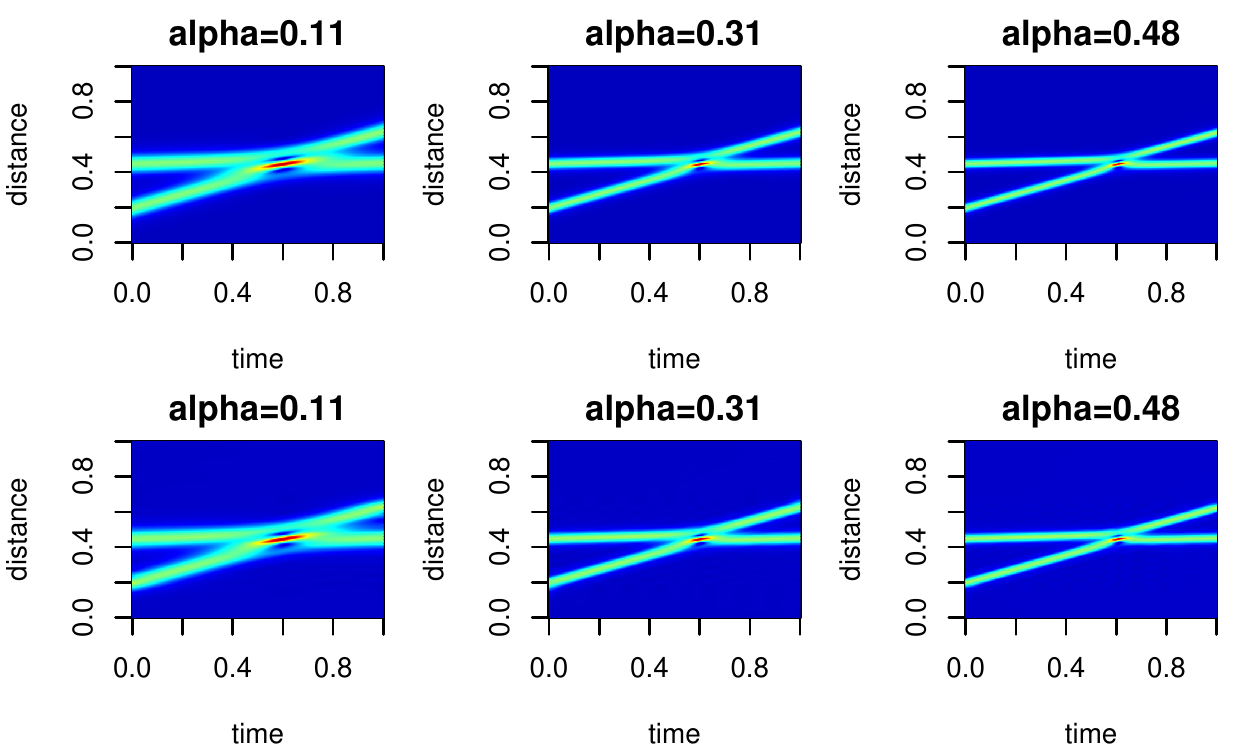}        
     \vspace{-8pt}
     \caption{Top: numerical solutions of the Schr\"{o}dinger equation that show the propagation of two solitons in time and space with different amplitudes; Bottom: reconstructed solutions using the POD bases predicted by the tree}
     \label{fig:schrodinger}
\end{figure}

We obtain the solutions of the Schr\"{o}dinger equation (\ref{eq:schrodinger}) for $t\in[0,40]$ and for 46 different values of amplitude $\alpha \in \Omega_{\alpha}=\{ 0.05, 0.051, 0.052\cdots, 0.49, 0.5\}$ . For any given $\alpha$, the obtained snapshot data are denoted by $\bm{D}(\alpha)$. 
As an illustration, the top row of Figure \ref{fig:schrodinger} shows the propagation of two solitons in time and space for amplitudes $\alpha=0.11$, $\alpha=0.31$ and $\alpha=0.48$. We see that the solution of the Schr\"{o}dinger equation depends on the amplitude parameter $\alpha$. 

In our experiment, we train the tree using the data generated from 13 amplitudes
\begin{equation} 
\begin{split}
\Omega_{\alpha}^{\text{train}} = \{ & 0.05, 0.09, 0.13, 0.17, 0.21, 0.25, 0.29, 0.33, 0.37, 0.41, 0.45, 0.49, 0.5\}
\end{split}
\end{equation}
Then, the tree is used to predict the POD bases for the remaining 33 amplitudes from the set $\Omega_{\alpha}^{\text{test}} = \Omega_{\alpha}\setminus\Omega_{\alpha}^{\text{train}}$. Finally, the error $e_{\alpha}$, $\alpha \in \Omega_{\alpha}$, is measured by the Frobenius norm:
\begin{equation}
e_{\alpha} = || \bm{D}(\alpha) - \hat{\bm{\Phi}}_{\alpha} \hat{\bm{\Phi}}_{\alpha}^T\bm{D}(\alpha) ||_F^2, \quad\quad \text{for } \alpha \in \Omega_{\alpha}^{\text{test}}
\end{equation}
where $\hat{\bm{\Phi}}_{\alpha}$ is the predicted POD basis by the tree for a given $\alpha \in \Omega_{\alpha}^{\text{test}}$. 

\begin{figure}[h!]
     \centering
     \begin{subfigure}[b]{0.48\textwidth}
         \centering
         \includegraphics[width=\textwidth]{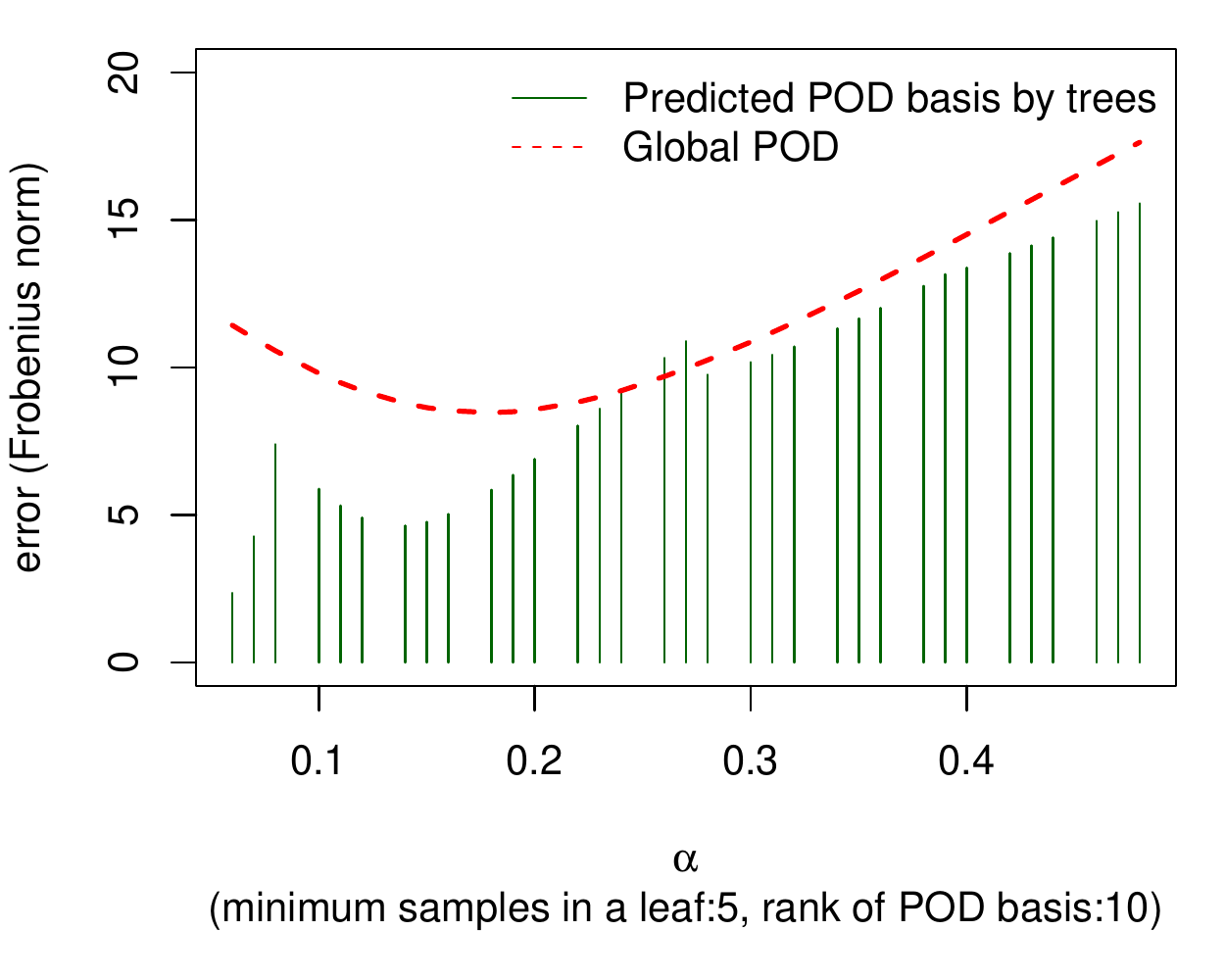}
         \caption{}
     \end{subfigure}
     \hfill
     \begin{subfigure}[b]{0.48\textwidth}
         \centering
         \includegraphics[width=\textwidth]{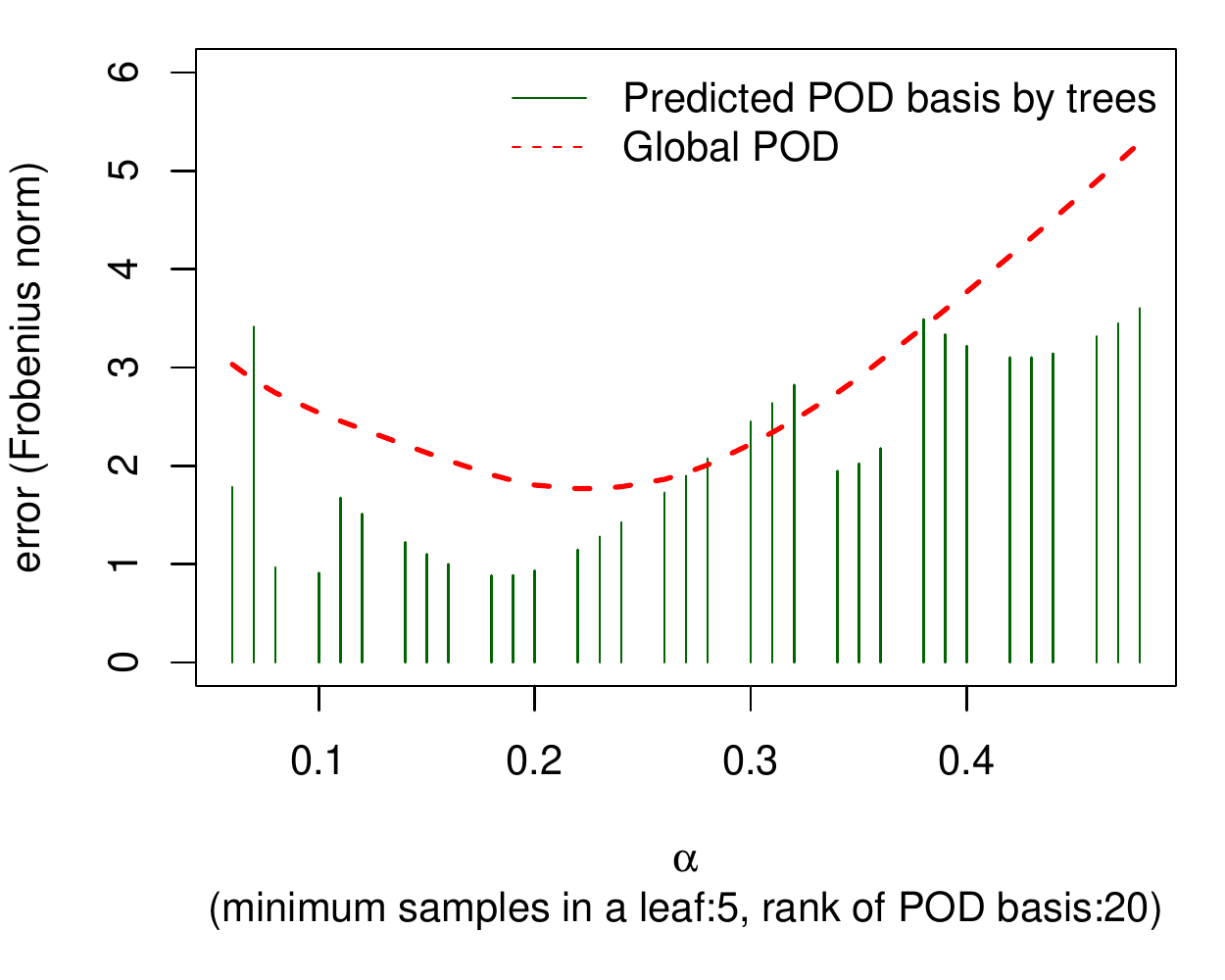}
         \caption{}
     \end{subfigure}
        \caption{The error, $e_{\alpha}$ for all $\alpha \in \Omega_{\alpha}$, based on both the predicted POD basis by the proposed trees and the global POD bases.}
        \label{fig:Schrodinger_result}
\end{figure}

As an illustration, the bottom row of Figure \ref{fig:schrodinger} shows the reconstructed solution, $\hat{\bm{\Phi}}_{\alpha} \hat{\bm{\Phi}}_{\alpha}^T\bm{D}(\alpha)$, at three amplitudes $\alpha=0.11$, $\alpha=0.31$ and $\alpha=0.48$, using the POD basis $\hat{\bm{\Phi}}_{\alpha}$ predicted by the tree.  
Figure \ref{fig:Schrodinger_result} shows the error, $e_{\alpha}$ for all $\alpha \in \Omega_{\alpha}$, based on both the POD basis predicted by the proposed trees and the global POD basis. In particular, we consider different ranks for the POD bases and let the minimum number of samples in a tree leaf to be 5. It is seen that, when the rank of the POD basis is 10, the prediction error of the proposed trees is lower than that using the global POD for 31 out of the 33 testing cases (Figure \ref{fig:Schrodinger_result}(a)). When the rank of the POD basis is increased to 20, the prediction error of the proposed trees is lower than that using the global POD for 27 out of the 33 testing cases (Figure \ref{fig:Schrodinger_result}(b)). We also note that, although increasing the rank of the global POD basis helps to reduce the error, using the POD bases predicted by the tree still yields much lower error for most of the parameter settings. 
\begin{figure}[h!]
     \centering
     \begin{subfigure}[b]{0.48\textwidth}
         \centering
         \includegraphics[width=\textwidth]{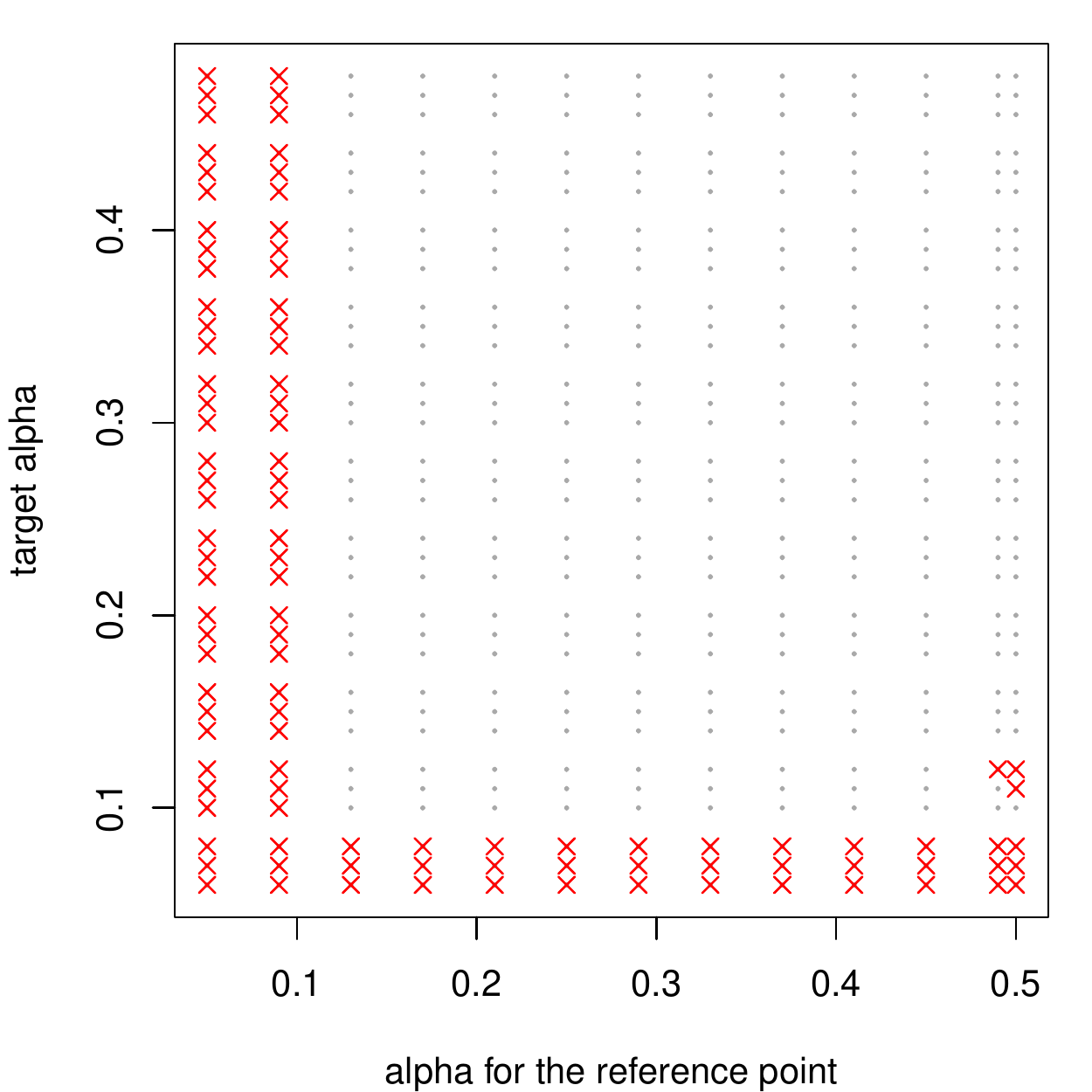}
         \caption{}
         \label{fig:three sin x}
     \end{subfigure}
      \hfill
       \begin{subfigure}[b]{0.48\textwidth}
         \centering
         \includegraphics[width=\textwidth]{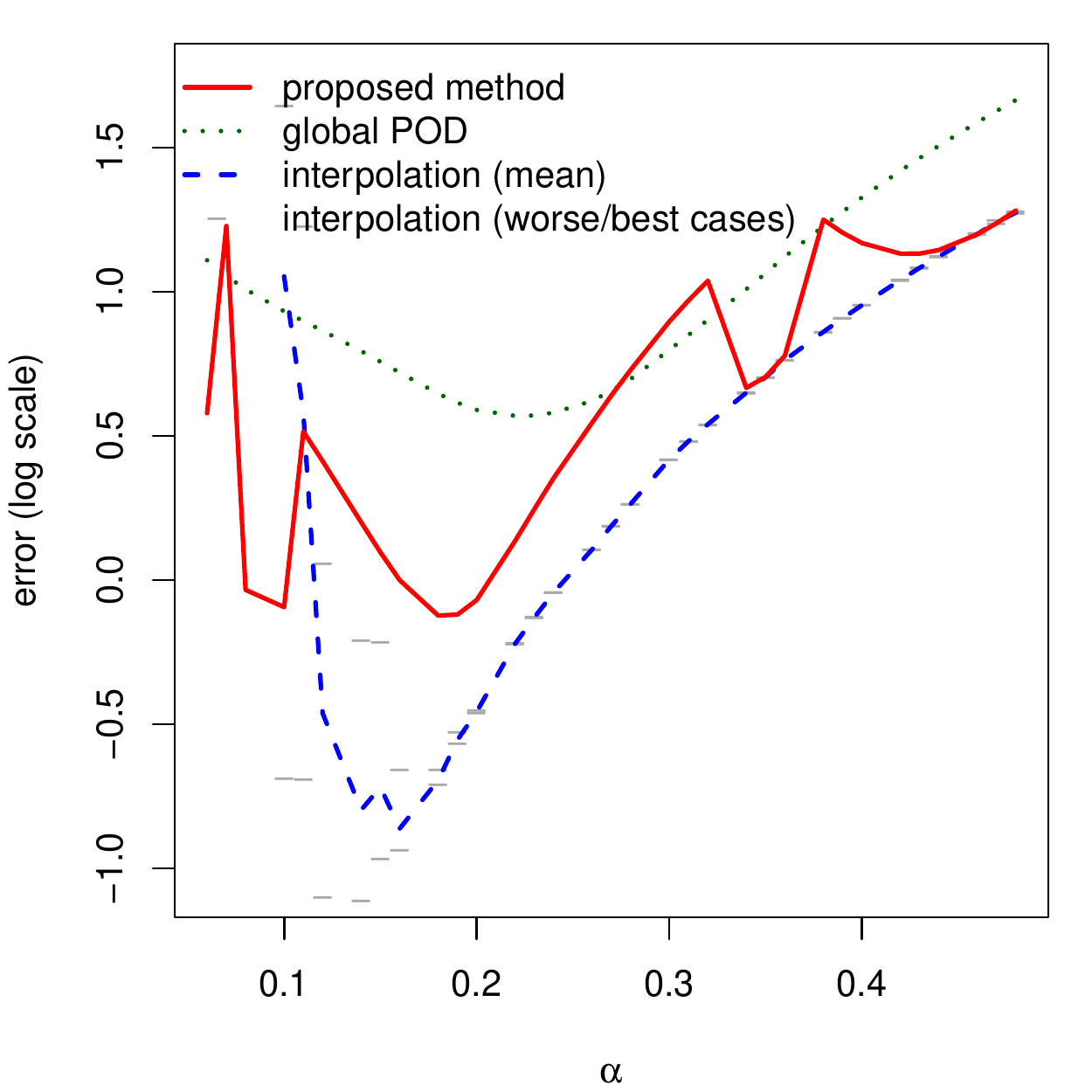}
         \caption{}
         \label{fig:three sin x}
     \end{subfigure}
        \caption{(a) Stability of the POD basis interpolation under the 33 parameter values of $\alpha$ for 13 different choices of the reference point (unstable interpolations are indicted by red crosses); (b) Comparison between the proposed tree-based method, global POD and the interpolation method}
        \label{fig:schrodinger_comparison_all}
\end{figure}

Similar to Example I, we further compare the proposed method to the existing interpolation method for POD basis via Grassmann manifolds. In this example, because there are 13 values for $\alpha$ in the training dataset, each of these 13 values can be used to establish the reference point for interpolation. Hence, we perform the POD basis interpolation under the 33 parameter values of $\alpha$ in the testing dataset for 13 times, and each time a particular value of $\alpha$ in the training set is used to establish the reference point. 

Figure \ref{fig:schrodinger_comparison_all}(a) firstly shows the stability of interpolation, while Figure  \ref{fig:schrodinger_comparison_all}(b) compares the prediction errors (in log scale) under different values of $\alpha$ between the proposed tree-based method, global POD and the interpolation method (with unstable interpolations being removed). Similar to Example I, some useful insights can be obtained: (i) The interpolations are unstable for values of $\alpha$ which are closer to the boundary of the range of $\alpha$; (ii) The stability of interpolation also depends on the choice of reference points. In this example, when the reference point is chosen corresponding to $\alpha=0.05$ and $\alpha=0.09$, the interpolation is unstable for all target values of $\alpha$; (iii) When the interpolation method is unstable or close to unstable for small target values of $\alpha$, the proposed tree-based method yields lower error, which suggests the potential hybrid use of the two methods.

\subsection{Example III: Thomas Young's double slit experiment}
In Example III, we consider the Thomas Young's Double Slit experiment governed by a partial differential equation 
\begin{equation}
u_{tt} - \Delta u = f_i 
\label{eq:wave}
\end{equation}

\vspace{-2pt}
\noindent with boundary conditions and initial values 
\begin{equation}
    \begin{split}
        u& = u(t)   \text {      on } \Gamma_{D}, \quad\quad n\cdot \nabla u  = 0  \text {      on }\Gamma_{N}\\
      u & = 0  \text{ for }   t=0, \quad\quad       \dot{u}  = 0   \text{ for }  t=0  
    \end{split}
\end{equation}

\begin{figure}[h!] 
     \centering
         \includegraphics[width=1\textwidth]{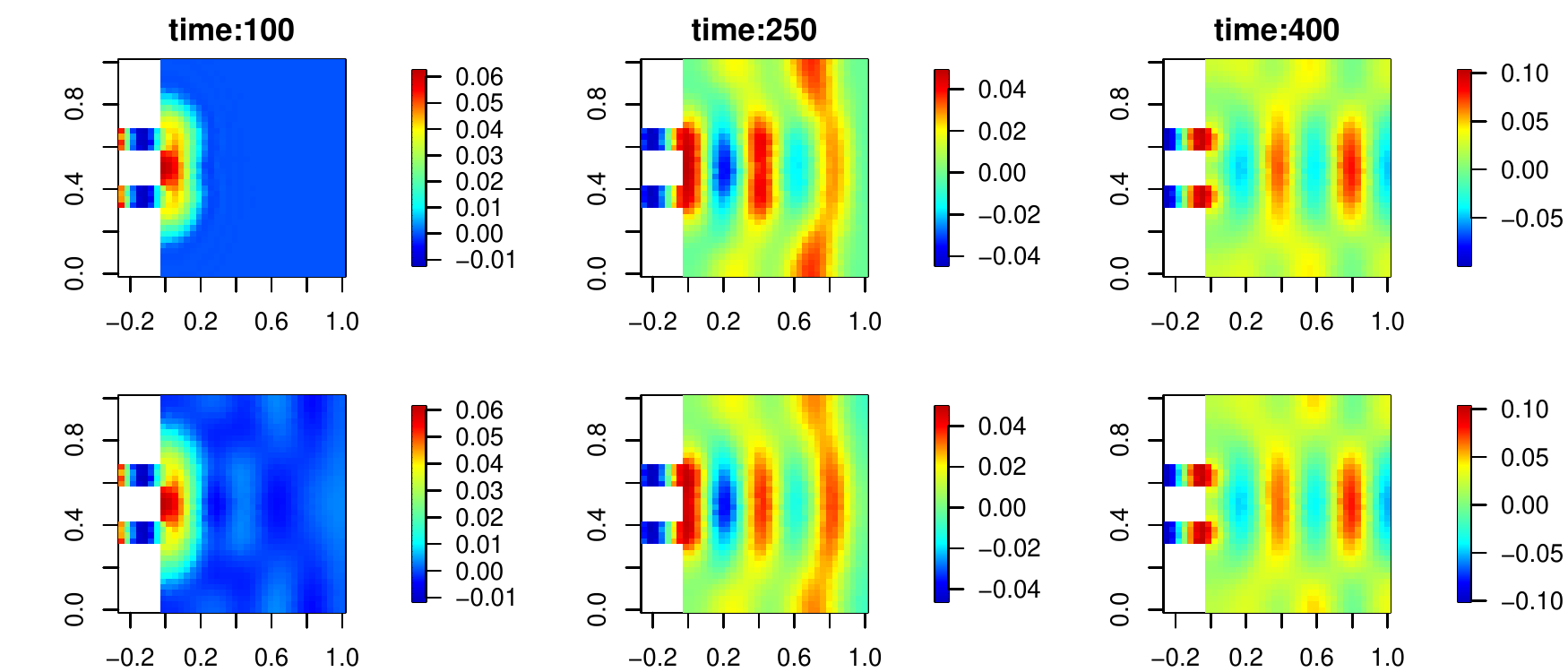}        
     \vspace{-8pt}
     \caption{Top: the solution of (\ref{eq:wave}) at $\bm{\mu}=(96, 5)$ at times 100, 250 and 400; Bottom: reconstructed solutions using the POD bases predicted by the tree.}
     \label{fig:doubleslit}
\end{figure}

\vspace{-2pt}
\noindent where $u$ denotes the quantity of displacement. The wave equation describes wave propagation in a median such as a liquid and a gas. As shown in Figure \ref{fig:doubleslit}, the domain of interest consists of a square with two smaller rectangular strips added on one side. The Dirichlet boundary condition is imposed on the line segments $\Gamma_{D}=\{x:x_1=-0.25\}$, and the Neumann boundary condition is imposed on the rest of the boundaries. The source is given by $f_i(x_{\Gamma_D})=\mu_1 \sin(\mu_2 \pi t)$, where $\bm{\mu}=(\mu_1,\mu_2)$ contains the parameters of interest. 

The Finite Element method is used to solve (\ref{eq:wave}) at 36 different combinations of $\mu_1$ and $\mu_2$ from a mesh grid of $\Omega_{\bm{\mu}}=\{80,84,88,92,96,100\} \otimes \{3.0,3.4,3.8,4.2,4.6,5.0\}$. Here, a number of 5731 elements and 2968 nodes are defined on the domain. For any given $\bm{\mu}$, the obtained snapshot data are denoted by $\bm{D}(\bm{\mu})$. 
As an illustration, the top row of Figure \ref{fig:doubleslit} shows the solution of (\ref{eq:wave}) at $\bm{\mu}=(96, 5)$ at times 100, 250 and 400.

In our experiment, we train the tree using the data generated from 12 randomly selected parameter settings
\begin{equation} 
\begin{split}
\Omega_{\bm{\mu}}^{\text{train}} = \{ & (80,4.2),(80,3.0),(100,4.2),(92,4.6),(84,4.6),(88,3.4),(88,5.0),\\ &(96,3.8),(92,3.0),(100,4.6),(92,3.8),(96,3.4)\}
\end{split}
\end{equation}
Then, the tree is used to predict the POD bases for the remaining 24 conditions from the set $\Omega_{\bm{\mu}}^{\text{test}} = \Omega_{\bm{\mu}}\setminus\Omega_{\bm{\mu}}^{\text{train}}$. Finally, the error $e_{\bm{\mu}}$, $\alpha \in \Omega_{\bm{\mu}}$, is measured by the Frobenius norm:
\begin{equation}
e_{\bm{\mu}} = || \bm{D}(\bm{\mu}) - \hat{\bm{\Phi}}_{\bm{\mu}} \hat{\bm{\Phi}}_{\bm{\mu}}^T\bm{D}(\bm{\mu}) ||_F^2, \quad\quad \text{for } \bm{\mu} \in \Omega_{\bm{\mu}}^{\text{test}}
\end{equation}
where $\hat{\bm{\Phi}}_{\bm{\mu}}$ is the predicted POD basis by the tree for a given $\bm{\mu} \in \Omega_{\bm{\mu}}^{\text{test}}$. 

\begin{figure}[h!]
     \centering
     \begin{subfigure}[b]{0.49\textwidth}
         \centering
         \includegraphics[width=\textwidth]{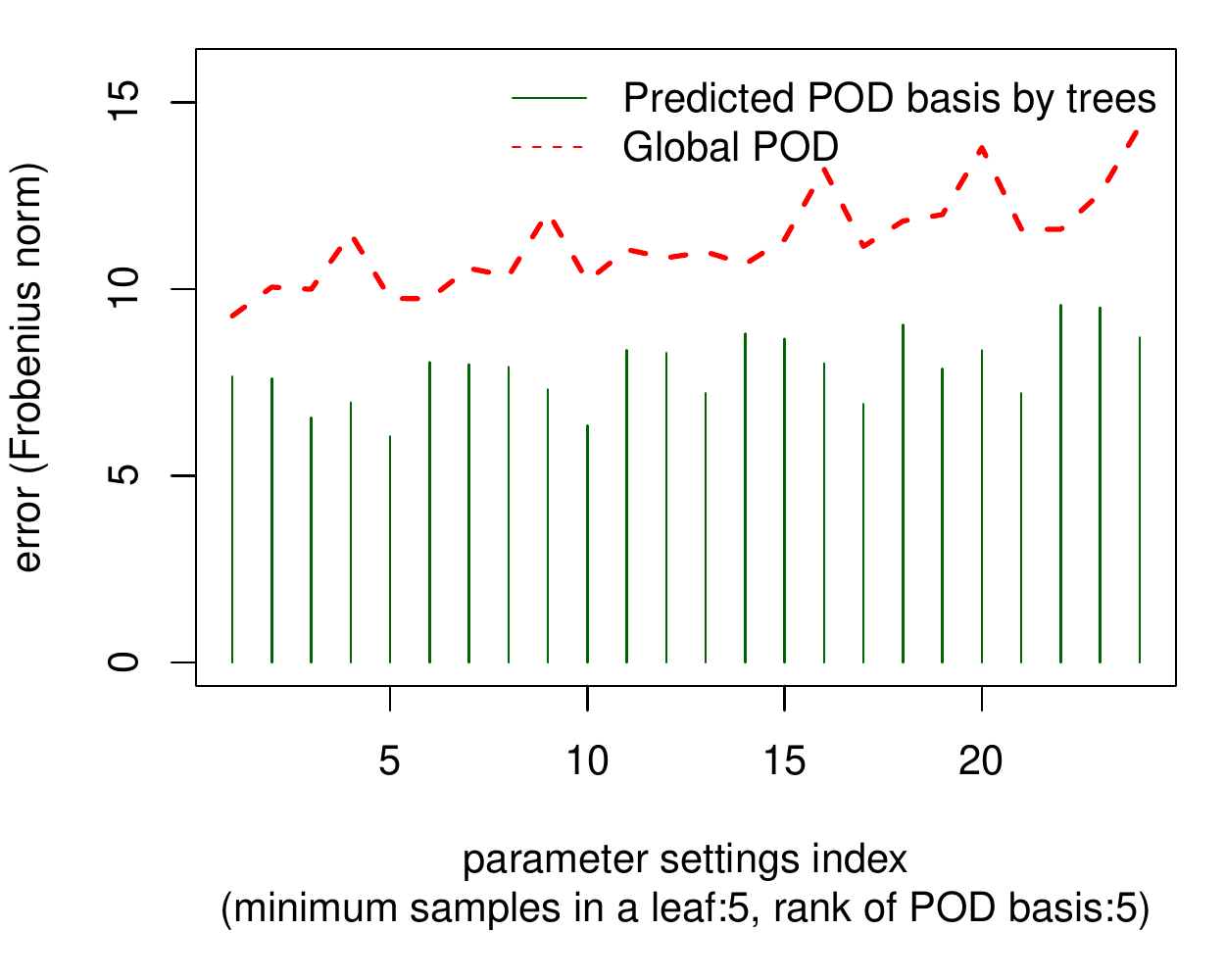}
         \caption{}
     \end{subfigure}
     \hfill
     \begin{subfigure}[b]{0.49\textwidth}
         \centering
         \includegraphics[width=\textwidth]{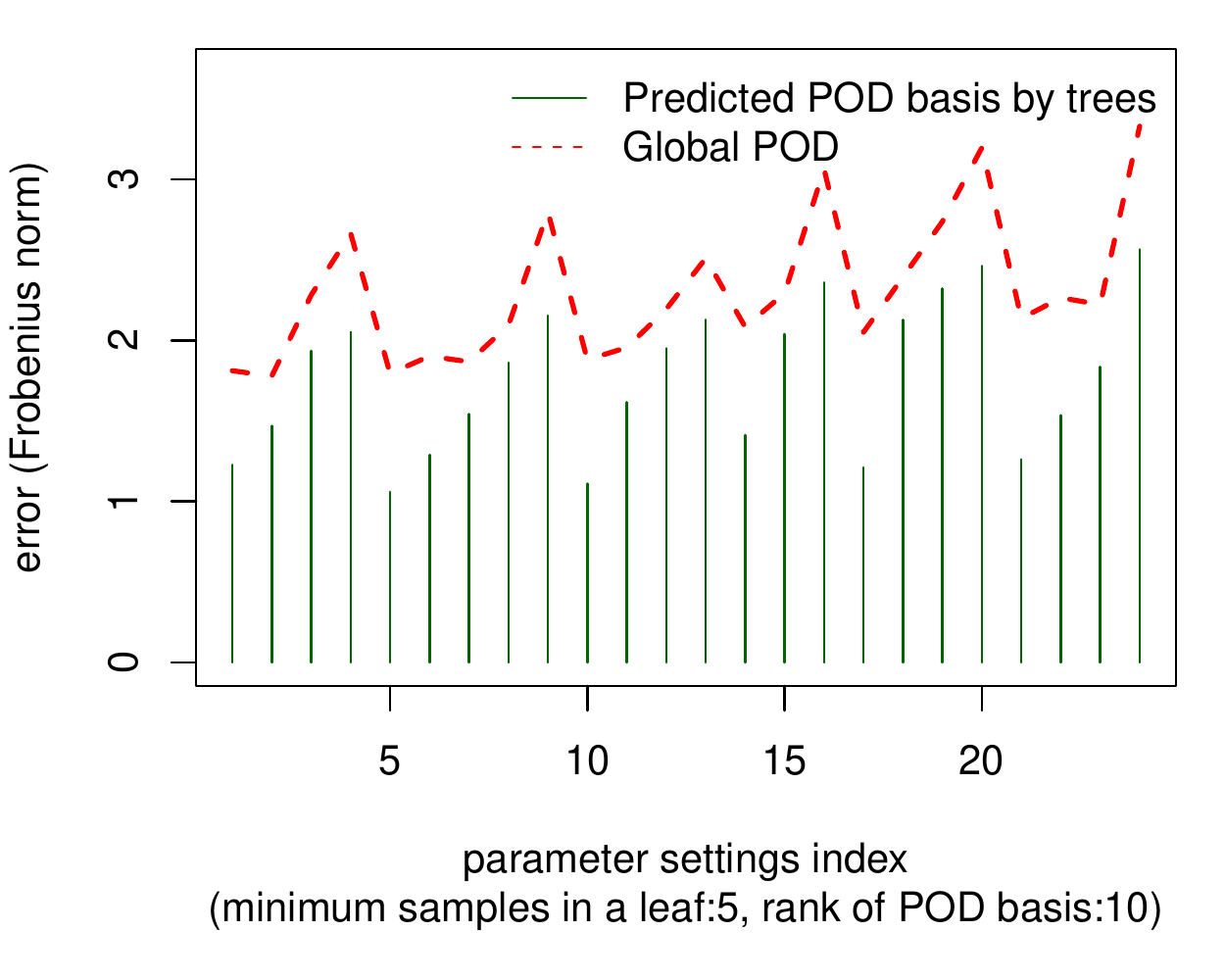}
         \caption{}
     \end{subfigure}
     \hfill
     \begin{subfigure}[b]{0.49\textwidth}
         \centering
         \includegraphics[width=\textwidth]{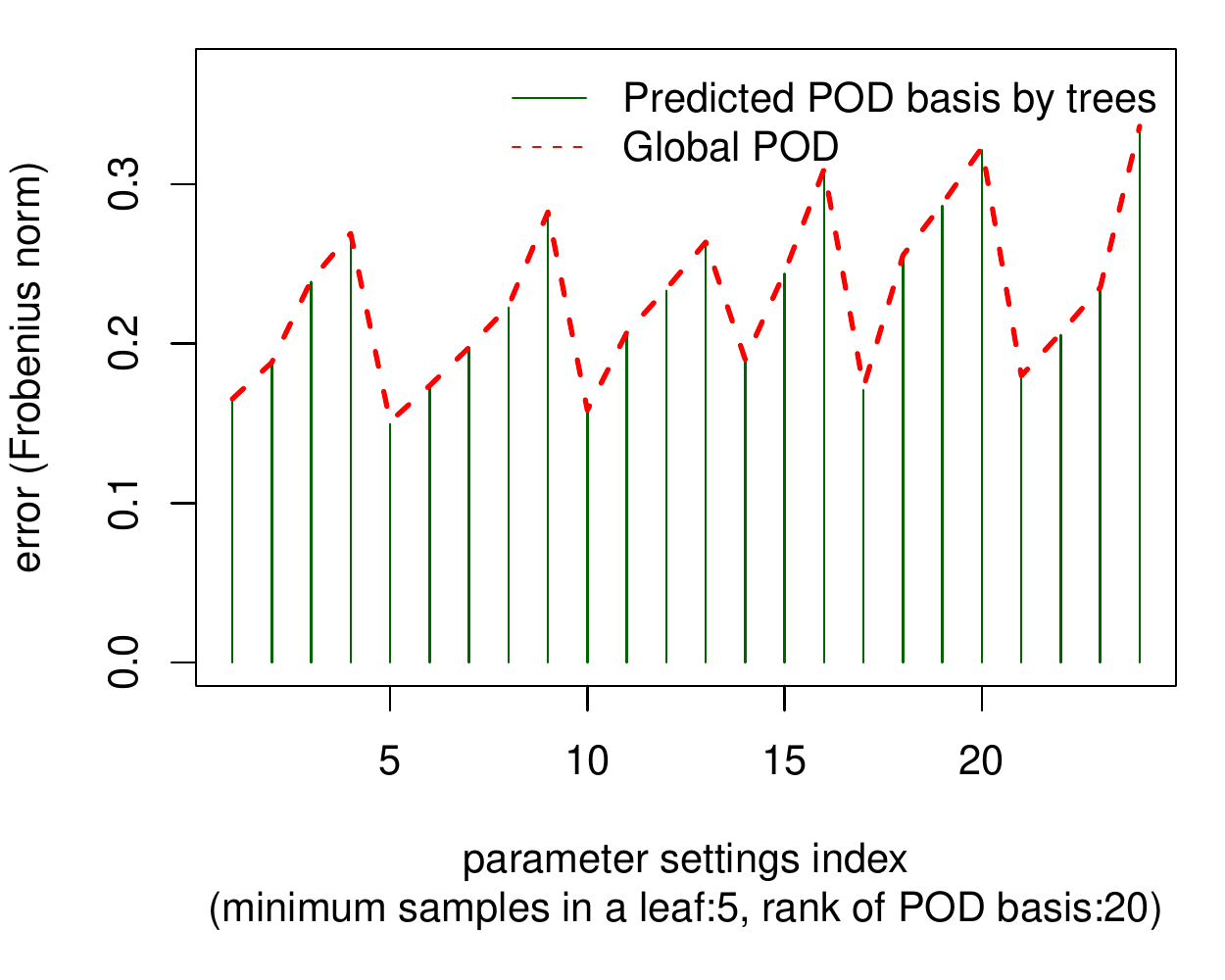}
         \caption{}
     \end{subfigure}
      \hfill
     \begin{subfigure}[b]{0.49\textwidth}
         \centering
         \includegraphics[width=\textwidth]{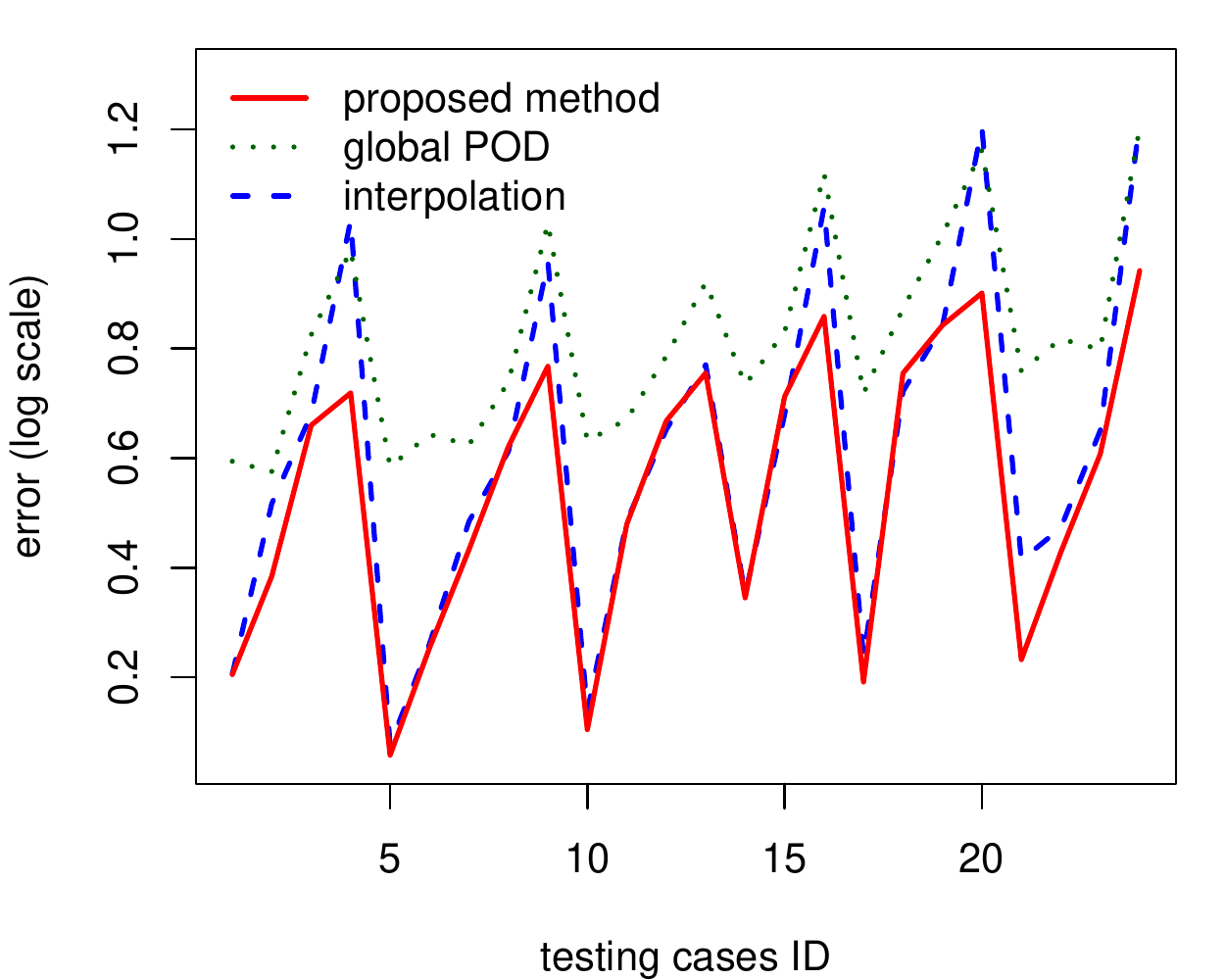}
         \caption{}
     \end{subfigure}
        \caption{(a $\sim$ c): The error, $e_{\bm{\mu}}$ for all $\bm{\mu} \in \Omega_{\bm{\mu}}$, based on both the predicted POD basis by the proposed trees and the global POD basis; (d) Comparison between the proposed tree-based method, global POD and the interpolation method.}
        \label{fig:doubleslit_result}
\end{figure}

As an illustration, the bottom row of Figure \ref{fig:doubleslit} shows the reconstructed solution using the POD basis $\hat{\bm{\Phi}}_{\bm{\mu}}$ predicted by the tree at times 100, 250 and 400 for $\bm{\mu}=(96,5)$.
Figure \ref{fig:doubleslit_result} shows the error, $e_{\bm{\mu}}$ for all $\bm{\mu} \in \Omega_{\bm{\mu}}^{\text{test}}$, based on both the predicted POD basis by the proposed trees and the global POD basis. As in previous examples, we consider different ranks for the POD bases and let the minimum number of samples in a tree leaf to be 5. It is seen that, when the ranks of the POD basis chosen to be 5 or 10, the prediction error of the proposed trees is consistently lower than that using the global POD for all 24 testing cases (Figure \ref{fig:doubleslit_result}(a) and (b)). When the rank of the POD basis is increased to 20, the prediction error of the proposed trees is approximately the same as using the global POD (Figure \ref{fig:doubleslit_result}(c)). This is mainly because the global POD basis with a rank of 20 can already capture 99.99\% energy of the data. This is consistent with our experiences that a high rank of the global POD basis is needed to reliably and accurately obtain a ROM at new parameter settings, which inevitably increase the computational cost. 

Similar to the previous examples, we further compare the proposed method to the existing interpolation method for POD basis via Grassmann manifolds. In this example, because there are respectively 12 and 24 parameter settings in the training and testing datasets, we perform the POD basis interpolation under the 24 parameter settings in the testing dataset for 12 times, and each time a particular setting in the training dataset is used to establish the reference point (the results show that all interpolations are stable for this example). 
Figure \ref{fig:doubleslit_result}(d) compares the prediction errors (in log scale) between the proposed tree-based method, global POD and the interpolation method. In this example, the interpolations turn out to be insensitive to the choice of reference point, thus the plot no longer includes the best/worst cases. We see that, both the interpolation method and the proposed method outperform the use of global POD bases, and the proposed tree-based method yields the lowest errors for most of the testing cases.

\subsection{Example IV: EBM additive manufacturing process}

In Example IV, we consider the heat transfer phenomenon during the Electron Beam Melting (EBM) additive manufacturing process of pure tungsten \citep{zhao2021enhancing}. The underlying governing equation of the transient heat transfer problem is given as follows:
\begin{equation} \label{eq:heat transfer}
    \rho c_p \frac{\partial \bm{T}}{\partial t}=\nabla \cdot (k \nabla T) + Q(\bm{x},t) \quad \text{in} \quad \Omega
\end{equation}
with the initial and boundary conditions:
\begin{subequations}
    \begin{alignat}{2}
        \bm{T}(\bm{x},0)& =\bm{T}_0  \quad && \text { in } \Omega\\
        \bm{T} & = \bm{T}_W  \quad &&\text { on }\Gamma_{1}\\
      -k\nabla\bm{T}\cdot\bm{n} & = q_0 \quad && \text { on }  \Gamma_2  \\ 
      -k\nabla\bm{T}\cdot\bm{n} & = h(\bm{T}_a-\bm{T}) \quad && \text { on }  \Gamma_3  
    \end{alignat}
\end{subequations}

Here, $\rho$ is the material density, $c_p$ is the specific heat capacity, $\bm{T}$ is the temperature, $k$ is the thermal conductivity, $\bm{T}_0$ is the initial temperature distribution, $\bm{T}_a$ is the ambient temperature, $\bm{T}_W$ is the temperature in Dirichlet boundary $\Gamma_1$, $q_0$ is the heat fluxes on Neumann boundary $\Gamma_2$, $h$ is the convection coefficient, $\Gamma_3$ is the convection boundary, $\Omega$ is the space-time domain, $\bm{n}$ is the unit vector outward normal to the boundary, and the absorbed heat flux $Q$ is defined as a Goldak heat source \citep{an2021implementation}:
\begin{equation}
    \begin{split}
        Q(\bm{x},t)_{f} &= \frac{6\sqrt{3}f_{f}P}{abc_{f}\pi \sqrt{\pi}}e^{\Big(-\frac{3(x_1+v\cdot t)^2}{c_{f}^2}\Big)}e^{\Big(-\frac{3x_2^2}{a^2}\Big)}e^{\Big(-\frac{3x_3^2}{b^2}\Big)}\\
        Q(\bm{x},t)_{r} &= \frac{6\sqrt{3}f_{r}P}{abc_{r}\pi \sqrt{\pi}}e^{\Big(-\frac{3(x_1+v\cdot t)^2}{c_{r}^2}\Big)}e^{\Big(-\frac{3x_2^2}{a^2}\Big)}e^{\Big(-\frac{3x_3^2}{b^2}\Big)}\\
        Q(\bm{x},t) &= \left\{\begin{array}{ll}
Q(\bm{x},t)_{f}, \text{when}\ \  x_1+v\cdot t\geq0\\
Q(\bm{x},t)_{r}, \text{when}\ \  x_1+v\cdot t<0
\end{array}\right.
    \end{split}
\end{equation}
where the width $a$, the depth $b$, the real length $c_r$, and the front length $c_f$ are geometric parameters, $f_r = 2c_r/(c_r+c_f)$ is the heat deposited fractional factor in the rear, $f_f = 2c_f/(c_r+c_f)$ is the heat deposited fractional factor in the front, $P$ is the parameter representing energy input rate, $v$ is the scan speed ($100\text{mm}\cdot s^{-1}$), and $(x_1,x_2,x_3)$ define the initial position of heat source. 

\begin{figure}[h!] 
     \centering
         \includegraphics[width=0.9\textwidth]{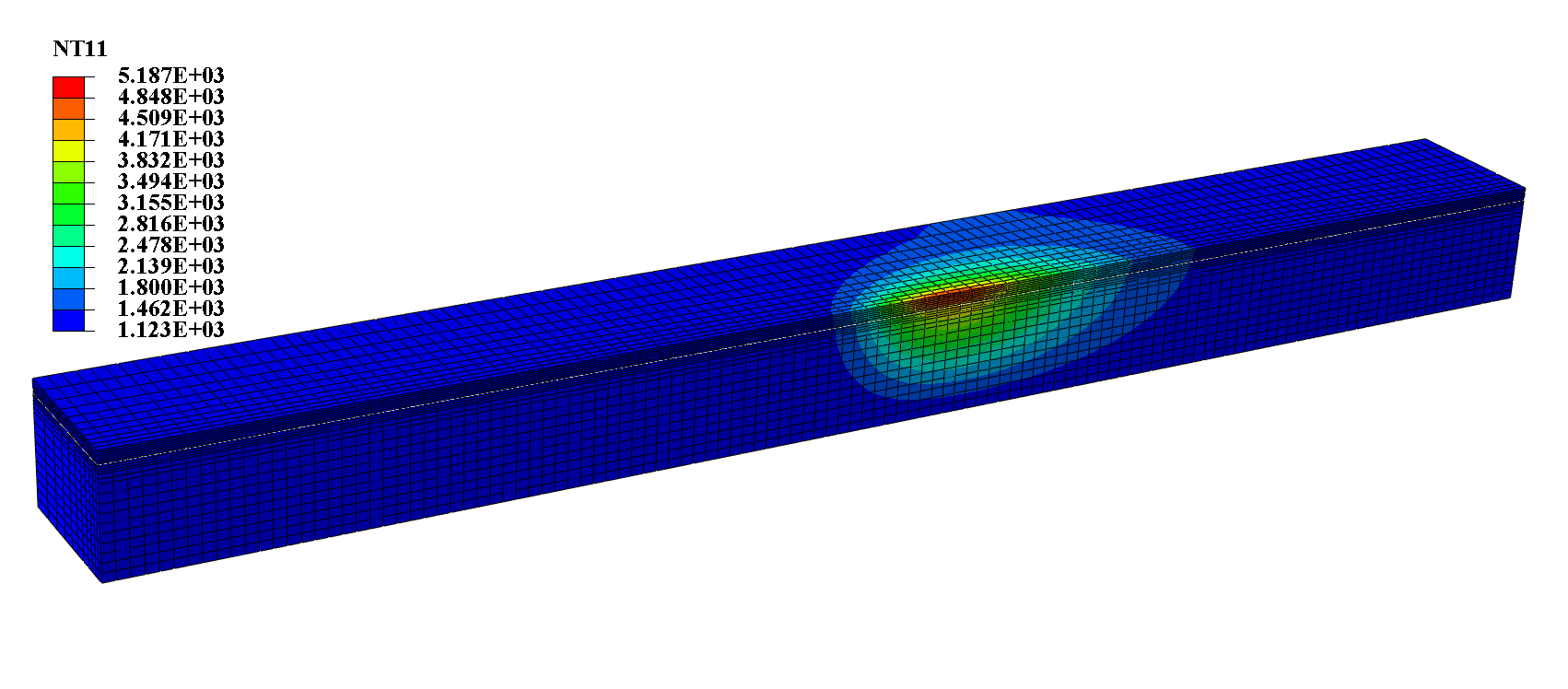}        
     \vspace{-16pt}
     \caption{A snapshot of the transient heat transfer process at time 1000 at an energy input rate of $P=579 \text{W}$}
     \label{fig:am}
\end{figure}
We solve this governing equation (\ref{eq:heat transfer}) by the Finite Element method for 46  energy input rates $P$ at $\Omega_{P} = \{5.79,5.8,5.81,...,6.23, 6.24\}\times 10^2 \text{W}$. Figure \ref{fig:am} shows the snapshot of the solution at time 1000 for $P=5.79\times 10^2 \text{W}$. The length of the meshed model is $8.4 \text{mm}$ and the width is $0.875 \text{mm}$. The height of the powder part is $0.07 \text{mm}$, and the height of the base part is $0.6 mm$. Only a half of the AM model is considered in FE model, because the printing area is located on the middle of the model. The melted area is discretized with dense mesh, while the surrounding area are meshed with larger elements.  

In our experiment, we train the tree using the data generated from 16 randomly selected parameter settings
\begin{equation} 
\begin{split}
\Omega_P^{\text{train}} = \{ & 5.79, 5.82, 5.85, 5.88, 5.91, 5.94, 5.97, 6.00,\\ &6.03, 6.06, 6.09, 6.12, 6.15, 6.18, 6.21, 6.24\}\times 10^2
\end{split}
\end{equation}
Then, the tree is used to predict the POD bases for the remaining 30 energy input rates from the set $\Omega_{P}^{\text{test}} = \Omega_P\setminus\Omega_{P}^{\text{train}}$. Finally, the error $e_{P}$, $\alpha \in \Omega_P$, is measured by the Frobenius norm:
\begin{equation}
e_P= || \bm{D}(P) - \hat{\bm{\Phi}}_{P} \hat{\bm{\Phi}}_{P}^T\bm{D}(P) ||_F^2, \quad\quad \text{for } P \in \Omega_{P}^{\text{test}}
\end{equation}
where $\hat{\bm{\Phi}}_{P}$ is the predicted POD basis by the tree for a given $P \in \Omega_{P}^{\text{test}}$. 

\begin{figure}[h!]
     \centering
     \begin{subfigure}[b]{0.49\textwidth}
         \centering
         \includegraphics[width=\textwidth]{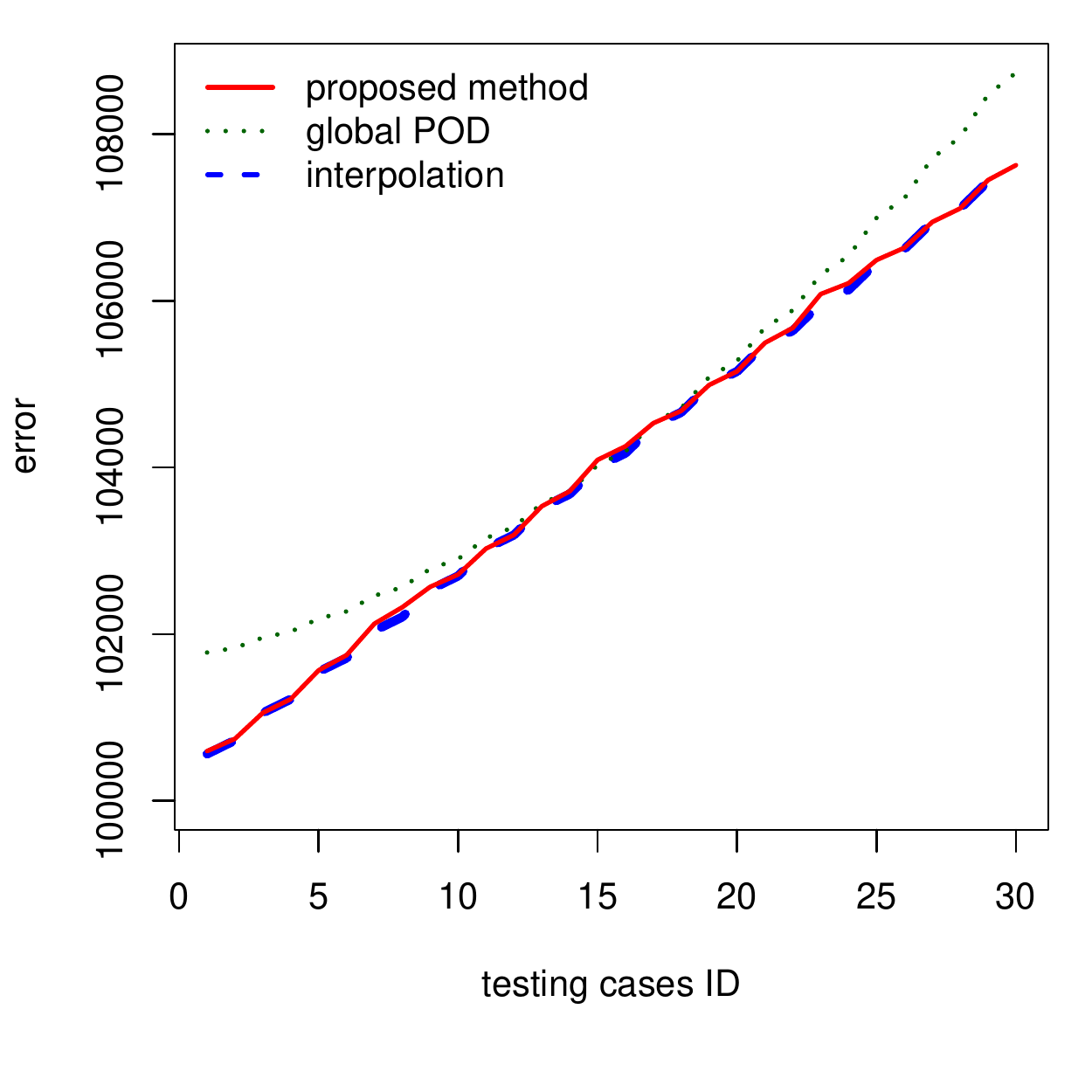}
         \caption{}
     \end{subfigure}
     \hfill
     \begin{subfigure}[b]{0.49\textwidth}
         \centering
         \includegraphics[width=\textwidth]{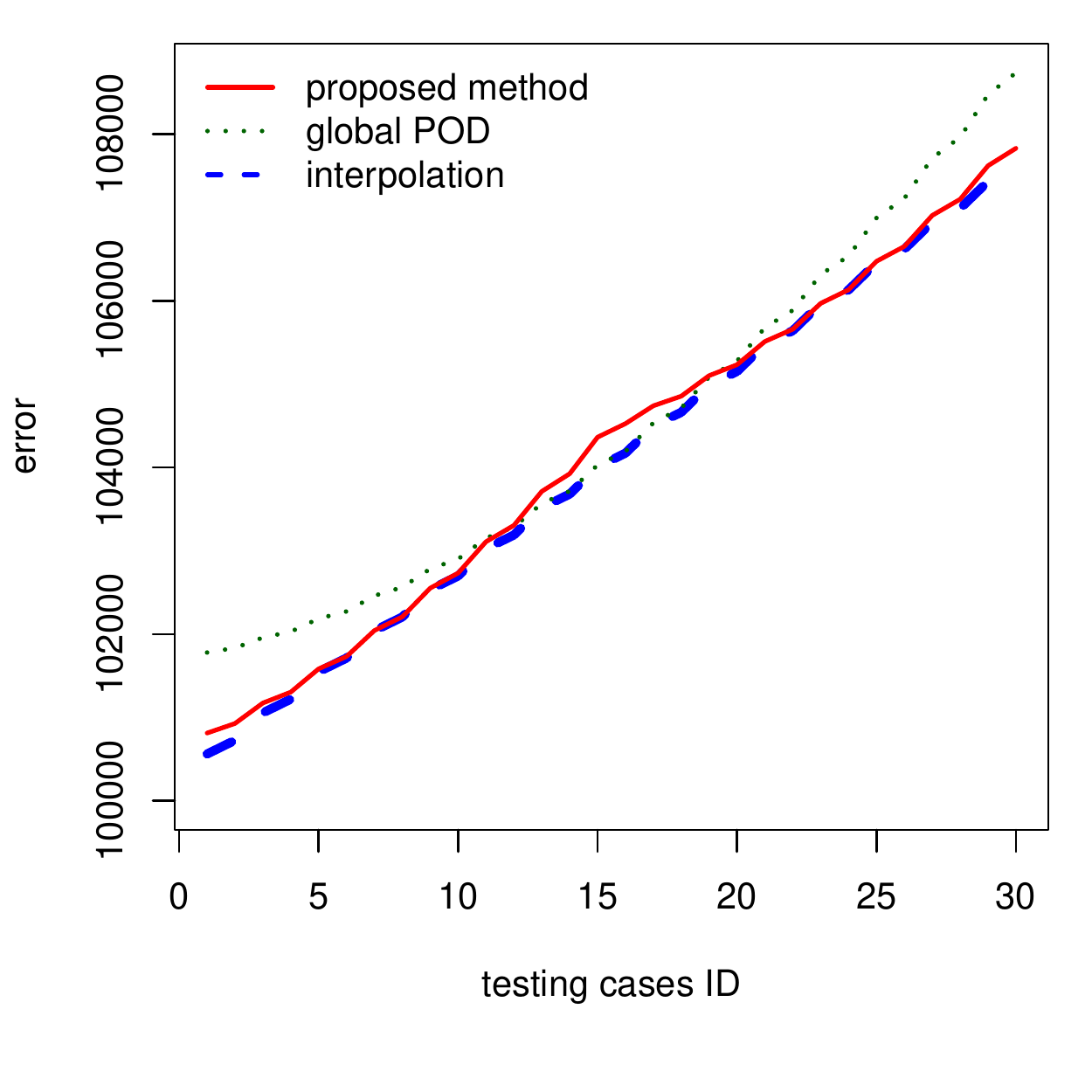}
         \caption{}
     \end{subfigure}
        \caption{Comparison between the proposed tree-based method, global POD and the interpolation method. The minimum samples in a tree leaf are set to 5 (panel (a)) and 10 (panel (b)). }
        \label{fig:am_result}
\end{figure}
Figure \ref{fig:am_result} compares the error, $e_{P}$ for all $P \in \Omega_{P}^{\text{test}}$, based on the predicted POD basis by the proposed trees, the global POD basis and the existing interpolation method for POD basis via Grassmann manifolds. As in previous examples, we let the minimum number of samples in a tree leaf to be 5 (Figure \ref{fig:am_result}(a)) and 10 (Figure \ref{fig:am_result}(b)), and set the rank for the POD basis to 10. In addition, because there are respectively 16 and 30 parameter settings in the training and testing datasets, we perform the POD basis interpolation for the 30 parameter settings in the testing dataset for 16 times, and each time a particular setting in the training dataset is used to establish the reference point (the results show that all interpolations are stable for this example). Figure \ref{fig:am_result}(a) clearly shows that the proposed method and the existing interpolation method have very close performance, and both methods outperform the use of global POD basis. We also note that the tree depth affects the performance of the proposed method, as expected for any regression trees. In this example, a deeper tree is required to accurately predict the POD basis at new parameter settings. In fact, for this particular example, there exists an almost linear relationship between the Euclidean distance (between any pair of energy input rates) and the Riemannian distance (between any two subspaces on the Grassmann manifold), making both the proposed and existing interpolation methods effective. More discussions are provided in the next section.

\subsection{Example V: additional discussions}
In the last example, we provide an example with some useful discussions about when the proposed approach is expected to perform well.
One key assumption behind the proposed tree-based method (as well as the existing POD basis interpolation method) is that similar parameter settings lead to similar POD basis. For two neighboring parameter settings, the subspaces spanned by the two POD bases are also expected to be close. However, \textit{when this critical assumption is not valid, it becomes difficult to learn the relationship between POD bases and parameters, if not impossible at all}.

\begin{figure}[h!]
     \centering
     \begin{subfigure}[b]{0.49\textwidth}
         \centering
         \includegraphics[width=\textwidth]{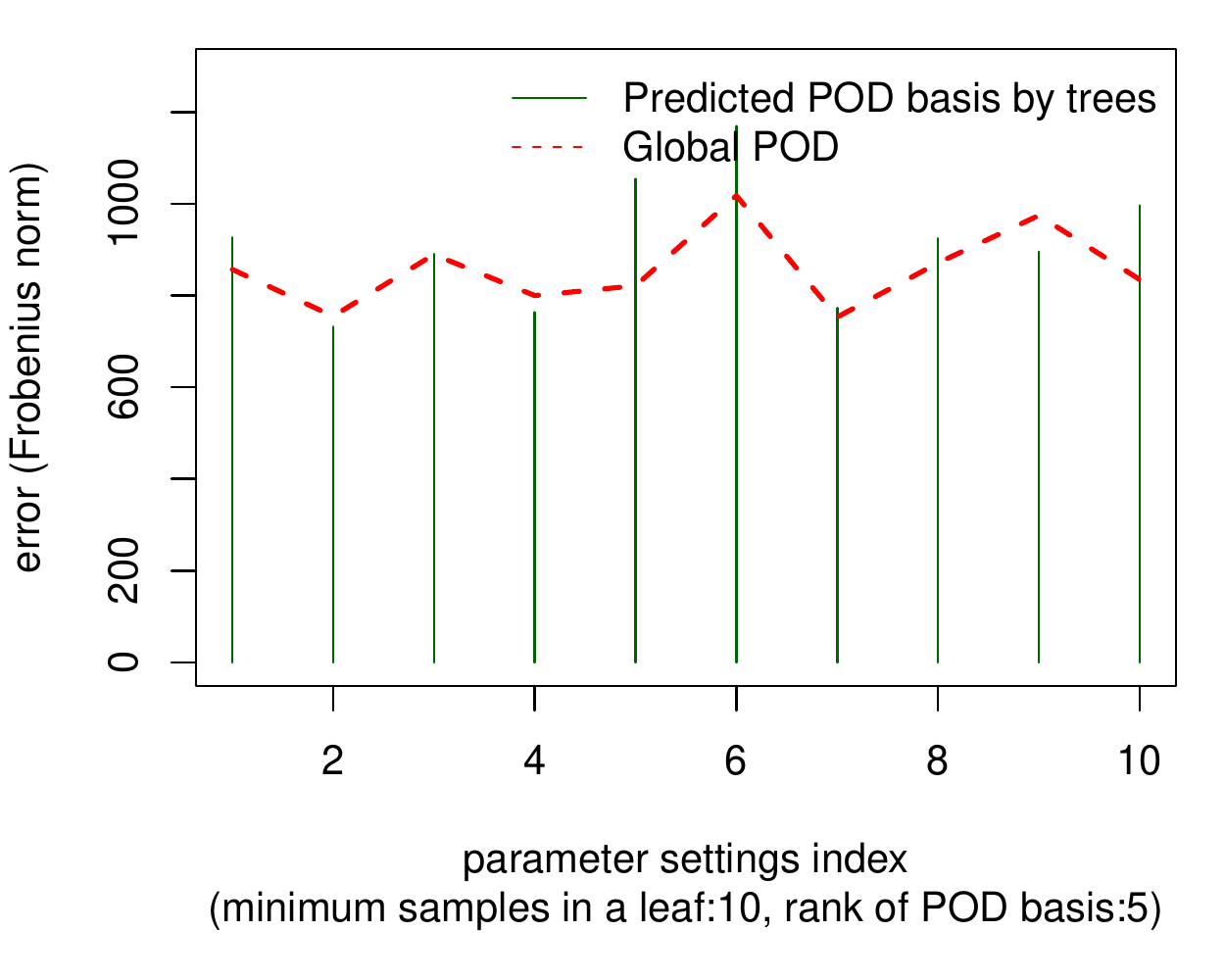}
         \caption{}
     \end{subfigure}
     \hfill
     \begin{subfigure}[b]{0.49\textwidth}
         \centering
         \includegraphics[width=\textwidth]{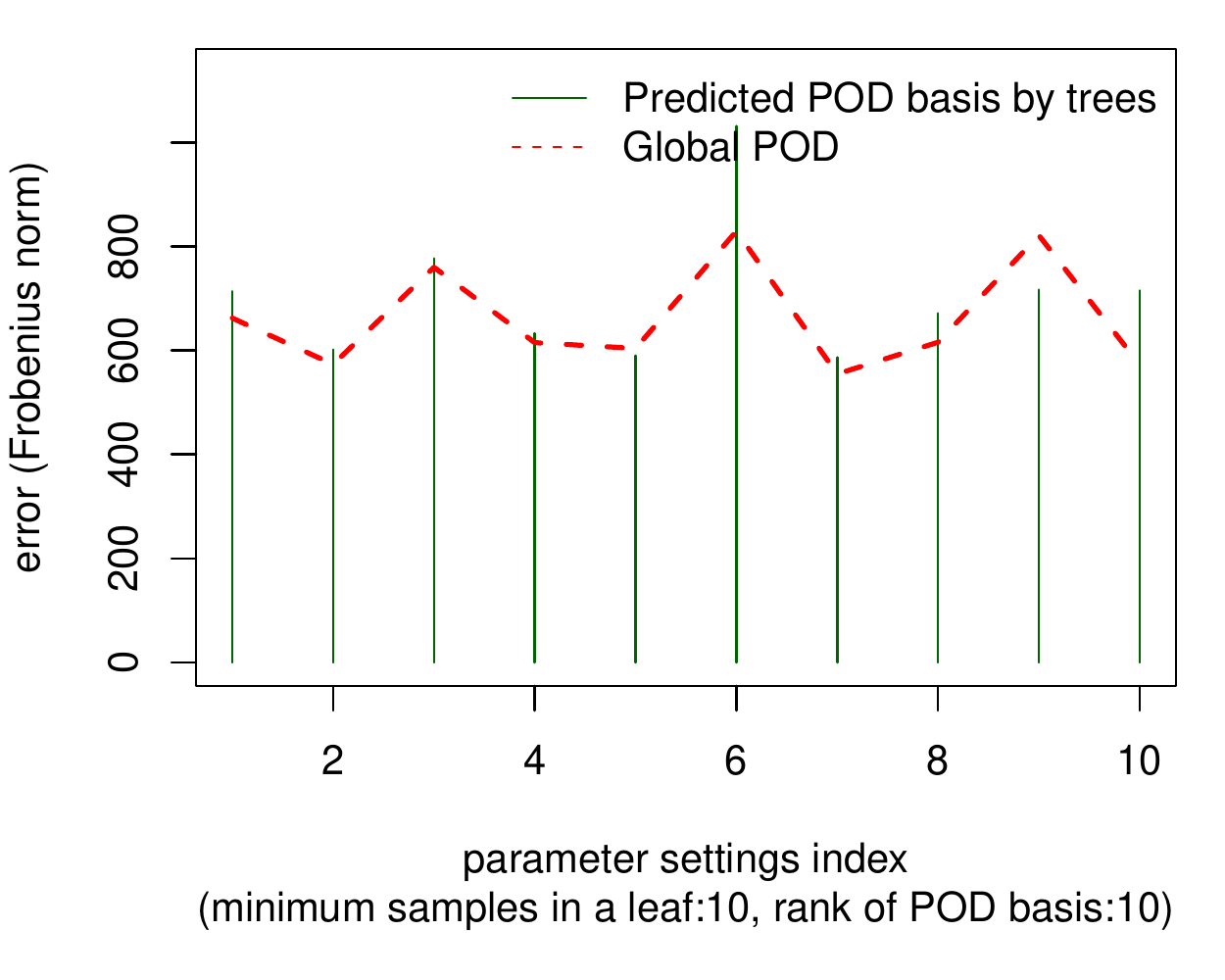}
         \caption{}
     \end{subfigure}
        \caption{Errors based on both the predicted POD basis by the proposed trees and the global POD basis.}
        \label{fig:UAV_result}
\end{figure}

\begin{figure}[h!]
    \centering
     \begin{subfigure}[b]{0.49\textwidth}
         \centering
         \includegraphics[width=\textwidth]{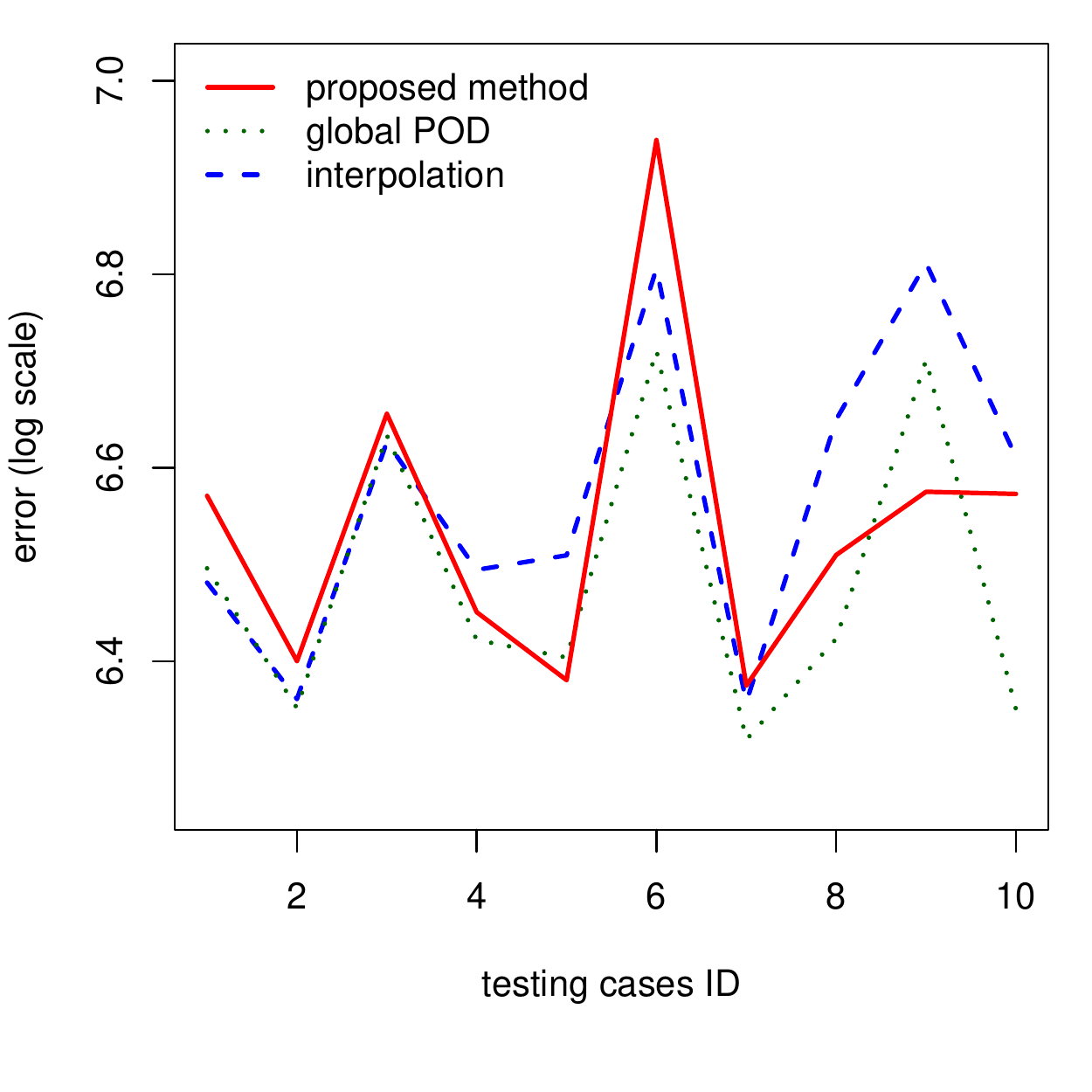}
         \caption{}
     \end{subfigure}
    \hfill
     \begin{subfigure}[b]{0.49\textwidth}
         \centering
         \includegraphics[width=\textwidth]{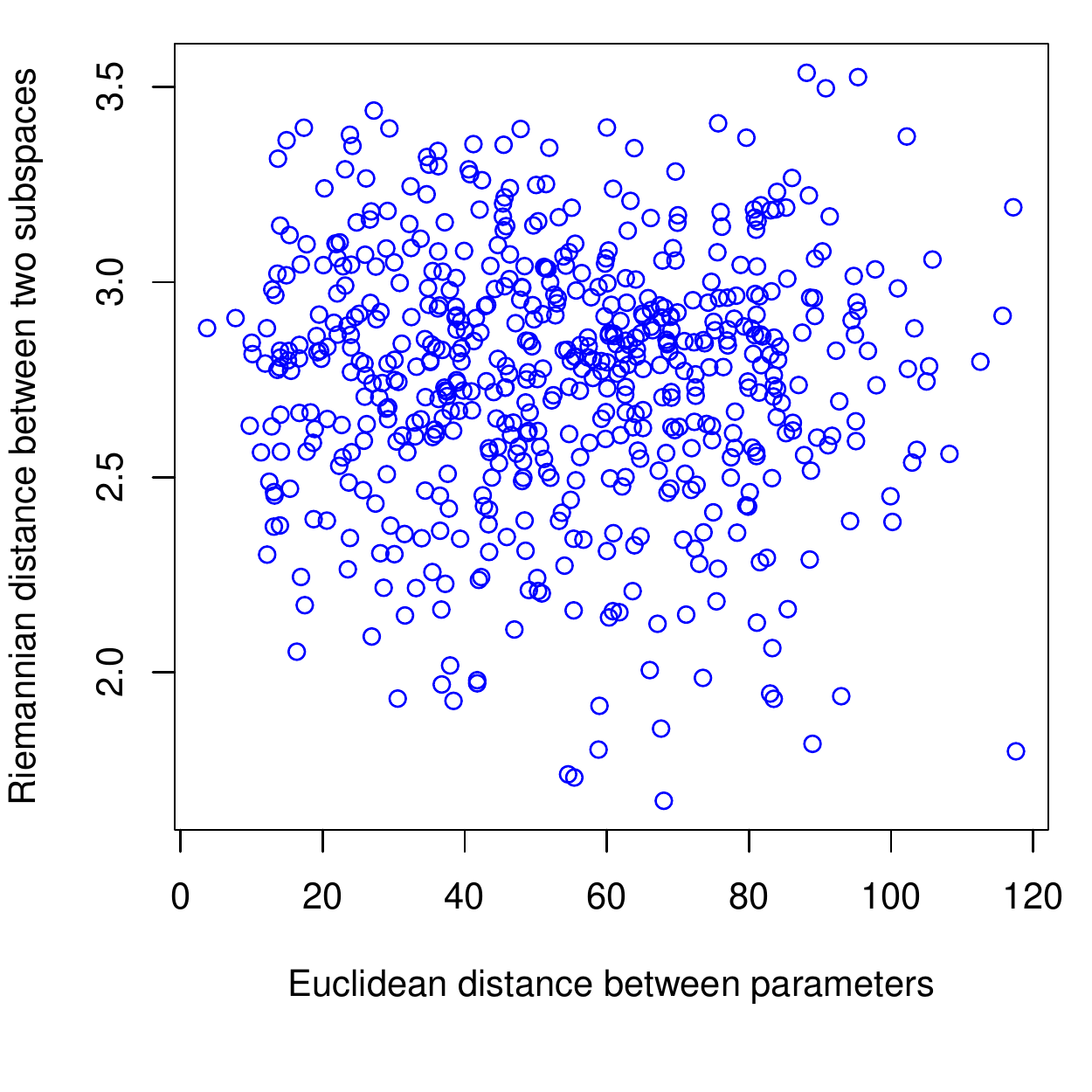}
         \caption{}
     \end{subfigure}
        \caption{(a) Comparison between the proposed tree-based method, global POD and the interpolation method. (b) Scatter plot between the Euclidean distance between two pairs of parameters and the Riemannian distance between two subspaces on the Grassmann manifold.}
        \label{fig:UAV_comparison_all}
\end{figure}

In Example V, we revisit the challenging nonlinear problem that involves the aircraft nose metal skin deformation process due to Unmanned Aerial Vehicle (UAV) collisions at different impact attitudes (i.e., pitch, yaw and roll degree); see Section \ref{sec:motivation}. A number of 35 snapshot datasets are obtained by FEA from the 35 collision conditions (i.e., combinations of pitch, yaw and roll degrees). 
To capture the relationship between collision attitudes and POD bases, we grow the proposed tree using the snapshot data from 25 randomly selected collision attitudes. Then, the tree is used to predict the POD bases for the remaining 10 conditions from the testing set, and the error is measured by the Frobenius norm. As in previous examples, we consider different ranks for the POD bases (5 and 10) and let the minimum number of samples in a tree leaf to be 10. The results are shown in Figure \ref{fig:UAV_result}. In addition, Figure \ref{fig:UAV_comparison_all}(a) shows the comparison between the proposed tree-based method, global POD and the interpolation method. 

\begin{figure}[h!]
     \centering
     \hfill
     \begin{subfigure}[b]{0.49\textwidth}
         \centering
         \includegraphics[width=\textwidth]{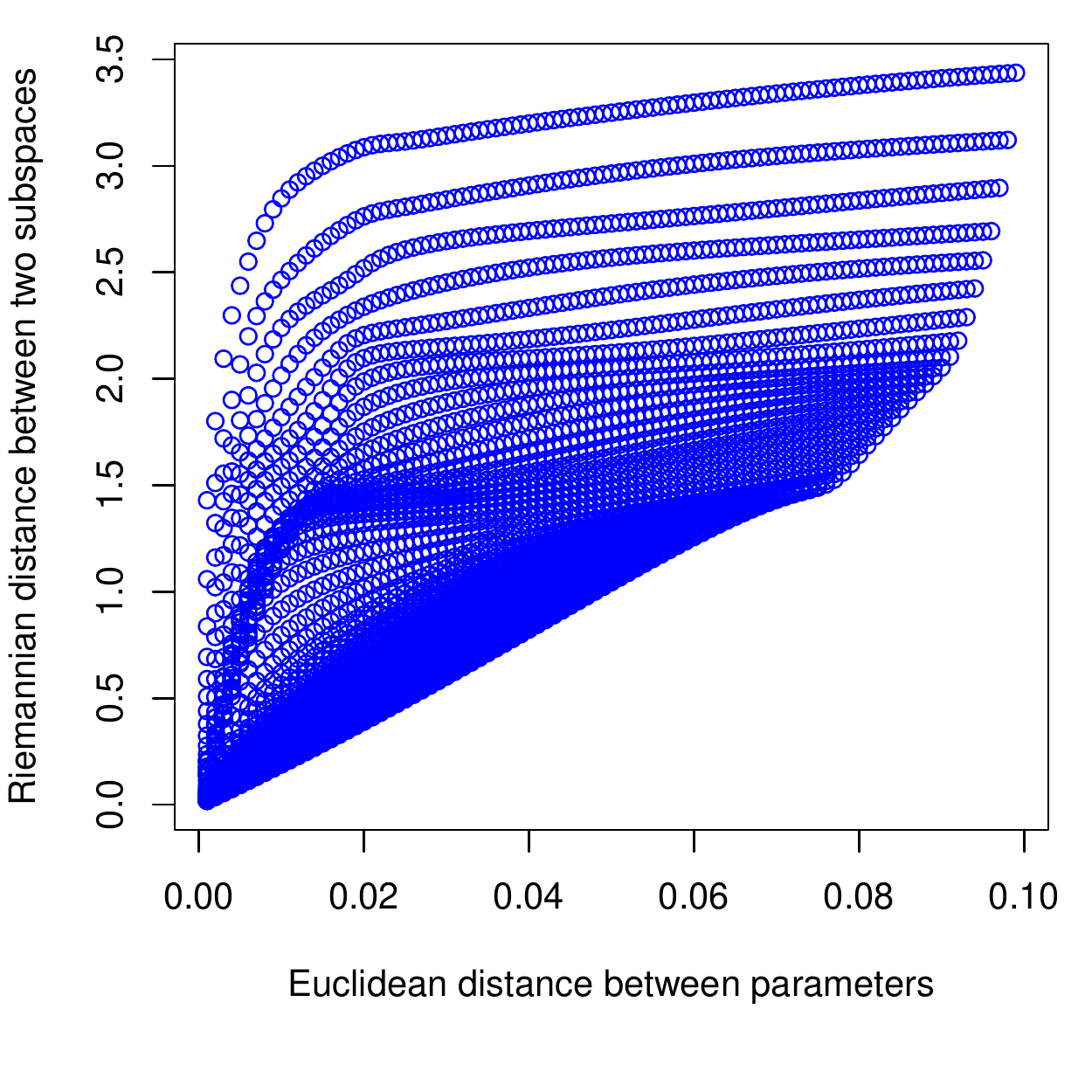}
         \caption{}
     \end{subfigure}
      \hfill
     \begin{subfigure}[b]{0.49\textwidth}
         \centering
         \includegraphics[width=\textwidth]{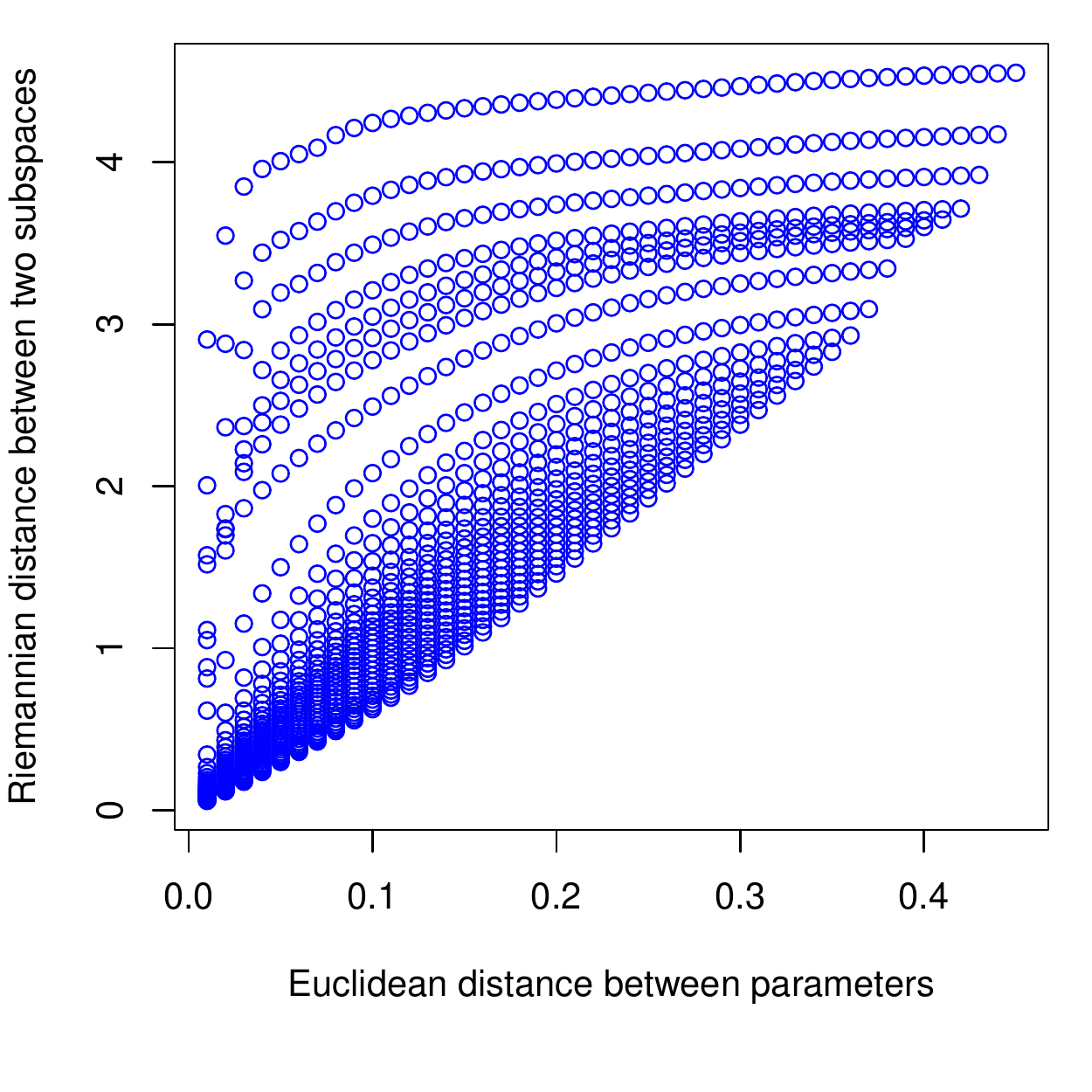}
         \caption{}
     \end{subfigure}
      \hfill
     \begin{subfigure}[b]{0.49\textwidth}
         \centering
         \includegraphics[width=\textwidth]{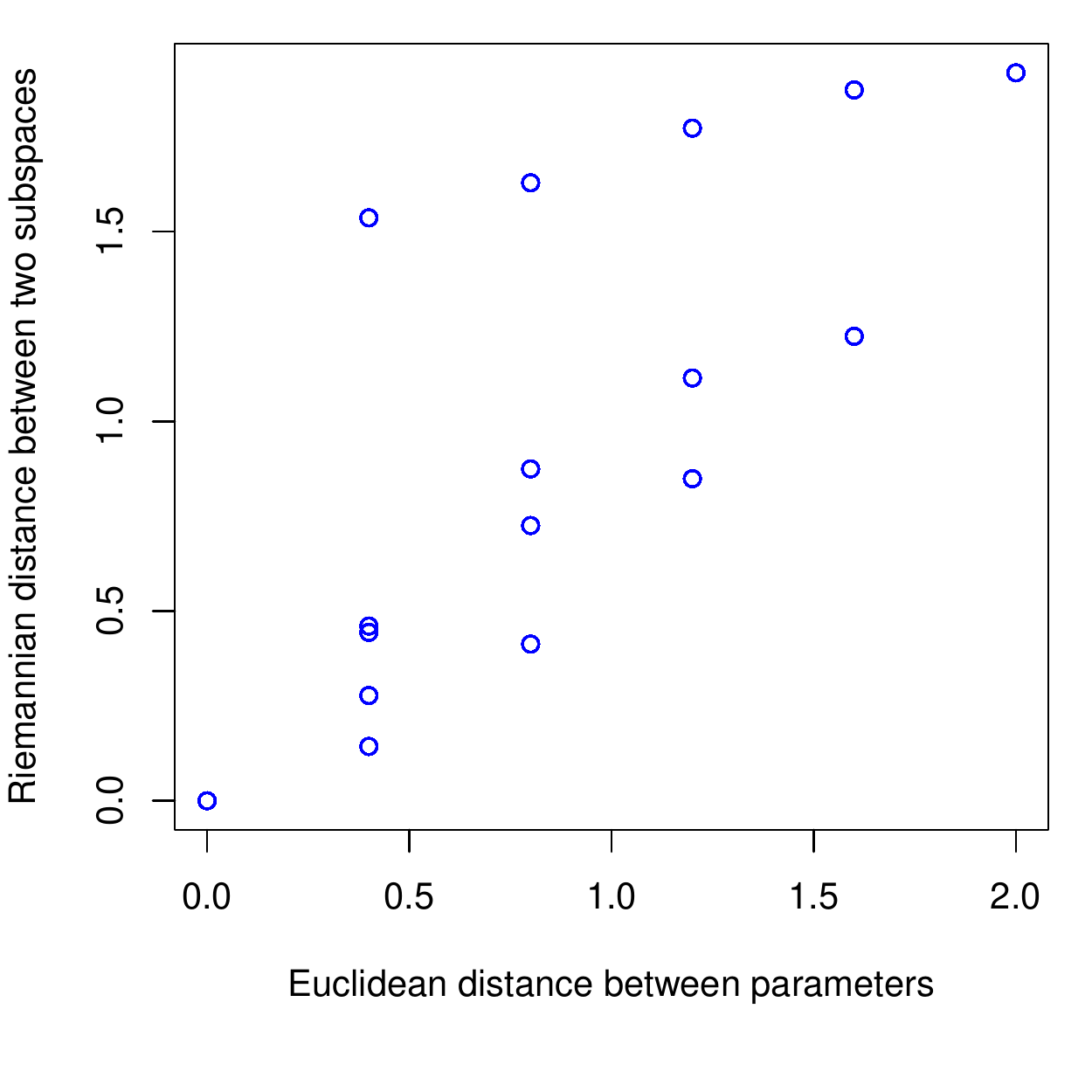}
         \caption{}
     \end{subfigure}
      \hfill
     \begin{subfigure}[b]{0.49\textwidth}
         \centering
         \includegraphics[width=\textwidth]{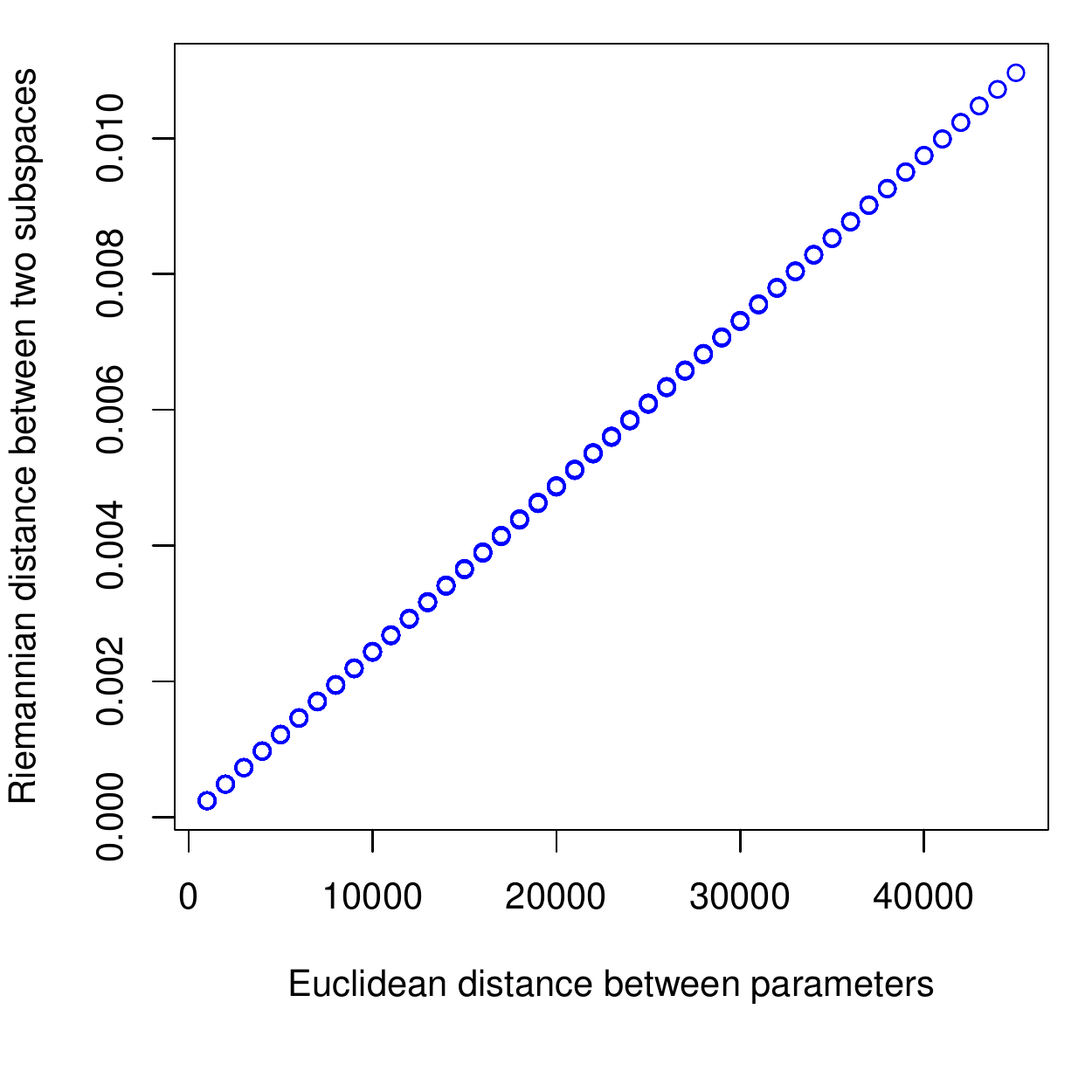}
         \caption{}
     \end{subfigure}
        \caption{Scatter plot between the Euclidean distance between two pairs of parameter settings and the Riemannian distance between two subspaces on the Grassmann manifold: (a) Example I: the heat equation (correlation: 0.71), (b) Example II: the Schr\"{o}dinger equation (correlation: 0.74), (c) Example III: the double slit experiment (correlation: 0.79), and (d) Example IV: the EBM additive manufacturing process (correlation: 0.99)}
        \label{fig:UAV_correlation}
\end{figure}

A quick examination of Figures \ref{fig:UAV_result} and \ref{fig:UAV_comparison_all}(a) suggests that the proposed tree-based method and the interpolation method do not outperform the use of global POD in this example. 
It is important to understand the reason behind this observation. Recall that the proposed tree-based method (as well as the existing POD basis interpolation methods) is based on a key assumption that similar parameter settings lead to similar POD bases. In other words, under similar collision attitudes, the subspaces spanned by the two POD bases should have a small Riemannian distance on the Grassmann manifold. 
To check this assumption, Figure \ref{fig:UAV_comparison_all}(b) shows the scatter plot between the Euclidean distance (between any pair of collision attitudes) and the Riemannian distance (between any two subspaces on the Grassmann manifold). Note that, we do not see a strong relationship between the Euclidean distance and Riemannian distance (in fact, the correlation coefficient is only -0.023). This observation suggests that although the POD bases change as collision attitudes change, similar collision attitudes do not necessarily lead to similar POD bases. Under this situation, we naturally do not expect the proposed statistical learning method, nor the existing interpolation method, to perform well. For comparison purposes, Figure \ref{fig:UAV_correlation} shows the scatter plot between the Euclidean distance and Riemannian distance for the previous Examples I, II, III and IV. For those examples, we clearly observe much stronger relationships between parameters and POD bases, and this explains why the proposed method and the existing interpolation approach outperform the use of global POD basis in the first four examples. In particular, for the EBM additive manufacturing example (i.e., Example IV), the correlation between the Euclidean distance of parameters and the Riemannian distance between subspaces reaches 0.99, which justifies why the interpolation method performs extremely well as shown in Figure \ref{fig:am_result}.

\section{Conclusions} \label{sec:conclusion}
This paper demonstrated the possibility and effectiveness of using statistical learning method (i.e., regression tress on Grassmann manifold) to learn a mapping between parameters and POD bases, and use the established mapping to predict POD bases for different parameter settings when constructing ROMs. The proposed tree is grown by repeatedly splitting the tree node to maximize the Riemannian distance between the two subspaces spanned by the predicted POD bases on the left and right daughter nodes. Five numerical examples were presented to demonstrate the capability of the proposed tree-based method. The comparison study showed that, when there exists a correlation between the Euclidean distance (between any pair of parameters) and the Riemannian distance (between any two subspaces on the Grassmann manifold), the proposed tree-based method as well as the existing POD basis interpolation method outperform the use of global POD, and thus better adapt ROMs for new parameters. The comparison study also showed that neither the proposed tree-based method nor the existing interpolation approach uniformly outperforms each other. Each method may outperform the other under certain situations, suggesting a potential hybrid use of these methods. 
Because regression tree is a relatively simple yet highly interpretable statistical model, more advanced statistical learning approaches will be investigated in the future to obtain more accurate predictions of POD basis.

\bibliography{references}

\end{document}